\documentclass{llncs}
\usepackage[latin1]{inputenc}

\usepackage{amsmath,amssymb}
\usepackage{latexsym,ifthen,epsfig,color}
\usepackage{float,enumerate,cite}
\usepackage{version}
\usepackage{tikz}

\newcommand{\join}{\text{\textcircled{{\footnotesize 1}}}}

\def\N{\mathbb{N}}
\def\cC{{\cal C}}
\def\cD{{\cal D}}
\def\cV{{\cal V}}
\def\cE{{\cal E}}
\def\cQ{{\cal Q}}

\newcommand{\NP}{\ensuremath{\mathbb{NP}}}

\newenvironment{myproof}[1][]{%
  \def\test{#1}%
  \ifx\test\empty%
  \begin{proof}%
  \else%
  \begin{proof}[#1]%
  \fi}{\qed\end{proof}}

\newenvironment{subproof}[1][]{%
  \def\test{#1}%
  \ifx\test\empty%
  \begin{proof}%
  \else%
  \begin{proof}[#1]%
  \fi}{\hspace*{\fill}{\tiny$\blacksquare$}\end{proof}}

\newtheorem{clai}{Claim}

\title{New Polynomial Cases of the Weighted Efficient Domination Problem}

\author{
Andreas Brandst\"adt\inst{1}
\and
Martin Milani\v c\inst{2}
\and
Ragnar Nevries\inst{1}
}

\institute{
Institut f\"ur Informatik, Universit\"at Rostock, D-18051 Rostock, Germany.\\
\email{\textrm{\{}andreas.brandstaedt,ragnar.nevries\textrm{\}}@uni-rostock.de}\\
\and
UP IAM and UP FAMNIT, University of Primorska, SI6000 Koper, Slovenia.\\
\email{martin.milanic@upr.si}
}

\includeversion{myproof}

\begin{document}

\maketitle

\begin{abstract}
Let $G$ be a finite undirected graph.
A vertex {\em dominates} itself and all its neighbors in $G$.
A vertex set $D$ is an {\em efficient dominating set} (\emph{e.d.}\ for short) of $G$ if every vertex of $G$ is dominated by exactly one vertex of $D$.
The \emph{Efficient Domination} (ED) problem, which asks for the existence of an e.d.\ in $G$, is known to be \NP-complete even for very restricted graph classes. 

In particular, the ED problem remains \NP-complete for $2P_3$-free graphs and thus for $P_7$-free graphs.
We show that the weighted version of the problem (abbreviated WED) is solvable in polynomial time on various subclasses of $2P_3$-free and $P_7$-free graphs, including $(P_2+P_4)$-free graphs, $P_5$-free graphs and other classes.

Furthermore, we show that a minimum weight e.d.\ consisting only of vertices of degree at most $2$ (if one exists) can be found in polynomial time.
This contrasts with our \NP-completeness result for the ED problem on planar bipartite graphs with maximum degree $3$.
\end{abstract}

\noindent{\small\textbf{Keywords}:
efficient domination;
$P_k$-free graphs;
polynomial time algorithm;
robust algorithm.
}

\section{Introduction}\label{sec:intro}

Packing and covering problems in graphs and hypergraphs and their relationships belong to the most fundamental topics in combinatorics and graph algorithms and have a wide spectrum of applications in computer science, operations research and many other fields. Packing problems ask for a maximum collection of objects which are not ``in conflict'', while covering problems ask for a minimum collection of objects which ``cover'' some or all others. A good example is the Exact Cover Problem
(X3C [SP2] in \cite{GarJoh1979}) asking for a subset ${\cal F'}$ of a set family ${\cal F}$ over a ground set, say $V$, covering every vertex in $V$ exactly once. It is well known that this problem is \NP-complete even for set families
containing only $3$-element sets (see \cite{GarJoh1979}) as shown by Karp \cite{Karp1972}.

The following variants of the domination problem are closely related to the Exact Cover Problem:
Let $G=(V,E)$ be a finite undirected graph.

A vertex $v$ {\em dominates} itself and its neighbors.
A vertex subset $D \subseteq V$ is an {\em efficient dominating set} ({\em e.d.} for short) of $G$ if every vertex of $G$ is dominated by exactly one vertex in $D$.
Obviously, $D$ is an e.d.\ of $G$ if and only if the subfamily of all closed neighborhoods of vertices in $D$ is an exact cover of the closed neighborhoods of $G$. Note that not every graph has an e.d.; the {\sc Efficient Dominating Set} (ED) problem asks for the existence of an e.d.\ in a given graph $G$.

The notion of efficient domination was introduced by Biggs \cite{Biggs1973} under the name {\em perfect code}.
In \cite{BanBarSla1988,BanBarHosSla1996}, among other results, it was shown that the ED problem is \NP-complete.
It is known that ED is \NP-complete even for bipartite graphs~\cite{YenLee1996}, chordal graphs~\cite{YenLee1996}, planar bipartite graphs~\cite{LuTan2002}, chordal bipartite graphs~\cite{LuTan2002}, and planar graphs with maximum degree $3$ \cite{Kratochvil91,FelHoo2000}.
Efficient dominating sets are also called {\em independent perfect dominating sets} in various papers, and a lot of work has been done on the ED problem which is motivated by various applications, among them coding theory and resource allocation in parallel computer networks; see, e.g., \cite{BanBarSla1988,BanBarHosSla1996,Biggs1973,ChaPanCoo1995,LiaLuTan1997,Lin1998,LivSto1988,LuTan2002,Milan2012,Yen1992,YenLee1996}.

In this paper, we will also consider the weighted version of the ED problem:
\medskip
\begin{center}
\fbox{\parbox{0.82\linewidth}{\noindent
{\sc Weighted Efficient Domination (WED)}\\[.8ex]
\begin{tabular*}{.9\textwidth}{rl}
{\em Instance:} & A graph $G=(V,E)$, vertex weights $\omega:V\to \mathbb{N}$.\\
{\em Task:} & Find an e.d.\ of minimum total weight,\\
&  or determine that $G$ contains no e.d.
\end{tabular*}
}}
\end{center}

\medskip

The WED (and consequently the ED) problem is solvable in polynomial time in trees~\cite{Yen1992}, cocomparability graphs~\cite{C97,ChaPanCoo1995}, split graphs~\cite{ChaLiu1993}, interval graphs~\cite{CL94,ChaPanCoo1995},
circular-arc graphs~\cite{CL94},
permutation graphs~\cite{LiaLuTan1997},
trapezoid graphs~\cite{LiaLuTan1997,Lin1998},
bipartite permutation graphs~\cite{LuTan2002},
distance-hereditary graphs~\cite{LuTan2002},
block graphs~\cite{YenLee1996} and
hereditary efficiently dominatable graphs~\cite{CouMakRot2000,Milan2012}.


For a set ${\cal F}$ of graphs, a graph $G$ is called {\em ${\cal F}$-free} if $G$ contains no induced subgraph from ${\cal F}$. For two graphs $F$ and $G$, we say that $G$ is $F$-free if it is $\{F\}$-free.
Let $P_k$ denote a chordless path with $k$ vertices, and let $P_i+P_j$ denote the disjoint union of $P_i$ and $P_j$.
We write $2P_i$ for $P_i+P_i$.
From the \NP-completeness result for chordal graphs in~\cite{YenLee1996} it follows that for $2P_3$-free graphs, the ED problem remains \NP-complete and thus, it is also \NP-complete for $P_7$-free graphs:



A set $M$ of edges in a graph $G$ is an \emph{efficient edge dominating set} of $G$ if and only if it is an e.d.\ in the line graph $L(G)$ of $G$.
These sets are also called \emph{dominating induced matchings} in some papers.
It is known that deciding if a given graph has an efficient edge dominating set is \NP-complete, see e.g.\ \cite{BraHunNev2010,BraMos2011,CarKorLoz2011,GriSlaSheHol1993,LuKoTan2002,LuTan1998}.
Hence, we have:
\begin{corollary}\label{EDlinegrNPc}
For line graphs, the ED problem is \NP-complete.
\end{corollary}

The graph $S_{1,2,2}$ consists of a chordless path $a,b,c,d,e$ and an additional vertex $f$ adjacent to $c$.
Since line graphs are claw-free and $S_{1,2,2}$ contains the claw as induced subgraph, the ED problem is \NP-complete on claw-free graphs and $S_{1,2,2}$-free graphs.

\medskip

In this paper, we present polynomial time algorithms for the WED problem for various subclasses of $2P_3$-free graphs as well as of $P_7$-free graphs and also sharpen one of the \NP-completeness results by showing that the ED problem remains \NP-complete for planar bipartite graphs of maximum degree~$3$.
Our algorithms are typically robust, in the sense that for the algorithm working on a given graph class ${\cal C}$, it is not necessary to recognize whether the input graph is in ${\cal C}$; the algorithm either solves the problem or finds out that the input graph is not in ${\cal C}$ \cite{Spinr2003}.
Contrary to the above \NP-completeness result on planar bipartite graphs of maximum degree $3$, we show that
it can be decided in polynomial time whether an input graph $G$ contains an e.d.\ $D$ containing only vertices of degree at most $2$ in $G$, and if this is the case, such an e.d.\ of minimum weight can also be found efficiently.


\begin{figure}[htb]
\centering
\scriptsize
\newlength{\DD}
\setlength{\DD}{.55cm}
\begin{tikzpicture}
\tikzstyle{new}=[fill=black!09]

\begin{scope}[every node/.style={draw,rectangle,align=center}]
\node[right] at (0,1) (p2p7) {$(P_2+P_7)$\\-free};
\path (p2p7.center |- 0,-1) node (p7) {$P_7$-free};
\node[right] at (.8,.3) (p3p3) {$2P_3$-free};
\node[right] at (.8,-.4) (s122) {$S_{1,2,2}$-free};

\path (s122.east) -- ++(0.5\DD,.4) coordinate (first);
\path (first |- 0,1) -- ++(0.5\DD,0) node[right] (p2p6) {$(P_2+P_6)$\\-free};
\path (p2p6.east) -- +(\DD,0) node[right] (p2p5) {$(P_2+P_5)$\\-free};

\path (0,0) -- (first) coordinate[pos=.5] (zero-first-center);
\path (p2p5.east |- 0,0) -- ++(0.5\DD,0) coordinate (second);
\path (first) -- (second) coordinate[pos=.5] (first-second-center);

\path (first-second-center |- 0,-1) node (p6) {$P_6$-free};

\path (second |- 0,1) -- +(0.5\DD,0) node[right,new] (p2p4) {$(P_2+P_4)$\\-free};
\path (p2p4.east) -- +(\DD,0) node[right,new] (p2p3) {$(P_2+P_3)$\\-free};

\path (p2p3.east |- 0,0) -- ++(0.5\DD,0) coordinate (third);
\path (second) -- (third) coordinate[pos=.5] (second-third-center);

\path (second-third-center) -- ++(.7,0) node[new] (p3p3s122) {$\{2P_3,S_{1,2,2}\}$\\-free};

\path (second-third-center |- 0,-1) -- +(0,0) node[left,new] (p6s122) {$\{P_6,S_{1,2,2}\}$\\-free};
\path (second-third-center |- 0,-1) -- +(\DD,0) node[right,new] (p5) {$P_5$-free};


\path (third) -- ++(0.5\DD,0) node[right,new] (p2p2) {$2P_2$\\-free};
\path (p2p2.south) -- ++(0,-\DD) node[below] (split) {split};

\path (p2p2.east |- 0,0) coordinate (fourth);
\path (third) -- (fourth) coordinate[pos=.5] (third-fourth-center);
\end{scope}

\draw[->]
  (p2p7) edge (p2p6)
  (p2p6) edge (p2p5)
  (p2p5) edge (p2p4)
  (p2p4) edge (p2p3)
  (p2p3) edge (p2p2)

  (p7) edge (p6)
  (p6) edge (p6s122)
  (p6s122) edge (p5)
  (p5) edge (p2p2)

  (p2p2) edge (split)

  (p2p7.230) edge (p7.145)
  (p2p6) edge (p6)
  (p2p5.south) -- (p5.north west)

  (p3p3) edge (p3p3s122)
  (p3p3s122) edge (p2p2)

  (s122) edge (p3p3s122)
  (s122) edge (p6s122)

  (p7.130) -- (p3p3.south west)
  (p3p3) edge (p2p3.south west);

\draw (first |- 0,1.5) -- (first |- 0,-1.5) -- ++(0,-1em);
\draw (second |- 0,1.5) -- (second |- 0,-1.5) -- ++(0,-1em);
\draw (third |- 0,1.5) -- (third |- 0,-1.5) -- ++(0,-1em);

\path (zero-first-center) -- ++(0,-1.5) -- ++(0,-.5em) node {\NP-complete};
\path (first-second-center) -- ++(0,-1.5) -- ++(0,-.5em) node {open};
\path (second-third-center) -- ++(0,-1.5) -- ++(0,-.5em) node {polynomial};
\path (third-fourth-center) -- ++(0,-1.5) -- ++(0,-.5em) node {linear};

\end{tikzpicture}
\caption{The complexity of the Efficient Dominating Set Problem on several graph classes.
The arrows denote graph class inclusions.
The results for the gray highlighted classes are introduced in this paper, and hold for the weighted case
of the problem.}
\label{fig:classes}
\end{figure}
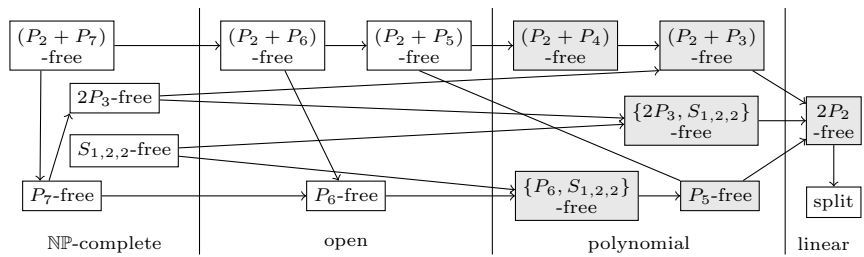

The paper is organized as follows:
Section \ref{2P2free} gives a linear time algorithm for this problem on $2P_2$-free graphs.
In Section \ref{P5freeED} we describe two ways for efficiently solving the ED problem on $P_5$-free graphs.
Sections \ref{P6S122free}, \ref{P3P3S122free} and \ref{P2P4free} contain polynomial time algorithms for $\left\{ P_6, S_{1,2,2} \right\}$-free graphs,  $\left\{ 2P_3, S_{1,2,2} \right\}$-free graphs, and $\left( P_2 + P_4 \right)$-free graphs, respectively.
Section \ref{2BED} gives a polynomial time algorithm that decides if a graph admits an e.d.\ consisting only of vertices of degree at most $2$. Finally, in Section \ref{EDNPc} we prove that the ED problem remains \NP-complete on planar bipartite graphs of maximum degree $3$.


\section{Basic Notions and Results}\label{sec:basicnotions}

All graphs considered in this paper will be finite, undirected and simple (i.e., without loops and multiple edges).
For a graph $G$, let $V$ denote its vertex set and $E$ its edge set; throughout this paper, let $|V|=n$ and $|E|=m$.
A graph is {\em nontrivial} if it has at least two vertices.
For a vertex $v \in V$, $N(v)=\{u \in V \mid uv \in E\}$ denotes its {\em open neighborhood}, and $N[v]:=\{v\} \cup N(v)$ denotes its {\em closed neighborhood}.
The {\em degree} of a vertex $x$ in a graph $G$ is $d(x):= |N(x)|$.
A vertex $v$ {\em sees} the vertices in $N(v)$ and {\em misses} all the others.
A vertex $u$ is {\em universal} for $G=(V,E)$ if $N[u]=V$.
Independent sets, complement graph, and connected components are defined as usual.

\medskip

Let $\delta_G(v,w)$ ($\delta(v,w)$ for short if $G$ is clear from the context) denote the distance between $v$ and $w$ in $G$.
The {\em square} of a graph $G = (V,E)$ is the graph $G^2 = (V,E^2)$ such that $uv\in E^2$ if and only if $\delta_G(u,v)\in\{1,2\}$.
In \cite{BraLeiRau2012,Leite2012,Milan2012}, the following relationship between the ED problem on a graph $G$ and the maximum weight independent set (MWIS) problem on $G^2$ is used:

\begin{lemma}\label{mainequived}
Let $G=(V,E)$ be a graph and $\omega(v):= |N[v]|$ a vertex weight function for $G$. Then the following are equivalent for any subset $D \subseteq V$:
\begin{enumerate}
\item[$(i)$] $D$ is an efficient dominating set in $G$.
\item[$(ii)$] $D$ is a minimum weight dominating set in $G$ with $\omega(D)=|V|$.
\item[$(iii)$] $D$ is a maximum weight independent set in $G^2$ with $\omega(D)=|V|$.
\end{enumerate}
\end{lemma}

Thus, the ED problem on a graph class ${\cal C}$ can be reduced to the MWIS problem on the squares of graphs in ${\cal C}$.
We will give an example for this reduction; in most cases, however, the direct way is more efficient.


Given a graph $G=(V,E)$ and a vertex $v \in V$, we define the {\em distance levels} $N_i(v) = \left\{ w \in V \mid \delta(v,w)=i \right\}$ for all $i \in \N$.
If $v$ is fixed, we denote $N_i(v)$ by $N_i$.

%
%

\subsection*{An Algorithmic Framework for the WED Problem}

In Sections~\ref{P5freeED}--
\ref{P2P4free} we will use the following algorithmic framework to solve the WED problem. For a specific graph class $\cC$, those sections concretize the subroutine \texttt{Robust-$\cC$-Best-Candidate-for-Vertex} used by the algorithm.

\medskip
\medskip
\hrule
\medskip
\noindent
\textbf{Algorithm:} \texttt{Robust-$\cC$-WED}\\
\textbf{Input:} A connected graph $G=(V,E)$ with vertex weights
$\omega:V\to \mathbb{N}$.\\
\textbf{Output:} One of the following: An e.d.\ $D$ of $G$ of minimum weight,
a proof that $G$ admits no e.d., or a proof that $G \not \in \cC$.
\begin{enumerate}
\item[(a)] Set $\cD := \emptyset$.
\item[(b)] For every vertex $v \in V$, do
\begin{enumerate}
\item[(b.1)] Determine the distance levels $N_1,N_2,\dots$ of $v$.
\item[(b.2)] Compute a set $D_v$ by calling \texttt{Robust-$\cC$-Best-Candidate-for-Vertex} for $v$ that is either an e.d.\ of $G$ with minimum weight over all e.d.s of $G$ containing $v$ (if $G$ admits an e.d.\ containing $v$), or not an e.d.\ at all.
If \texttt{Robust-$\cC$-Best-Candidate-for-Vertex} stops by proving that $G \not \in \cC$, \texttt{Stop}.
\item[(b.3)] Set $\cD := \cD \cup \{D_v\}$.
\end{enumerate}
\item[(c)] For every $D \in \cD$, check if $D$ is an e.d.\ of $G$ and calculate its weight.
\item[(d)] If $\cD$ contains no e.d.\ of $G$, \texttt{Stop}, otherwise, \texttt{Return a set $D \in \cD$} that is an e.d.\ of $G$ of minimum weight.
\end{enumerate}
\hrule

\medskip
The correctness of the algorithm can easily be seen.
Since determining the distance levels of a vertex $v$ can be done in linear time,
and checking if a vertex set is an e.d.\ can also be done in linear time, we get:

\begin{lemma}
If \texttt{Robust-$\cC$-Best-Candidate-for-Vertex} runs in time $O(t(n,m))$ for a graph class $\cC$
on an input graph with $n$ vertices and $m$ edges, then the WED problem is robustly
solvable on $\cC$ in time $O(n\cdot \max(n+m,t(n,m)))$.
\label{lma:abstractAlgo}
\end{lemma}


\section{The WED Problem for $2P_2$-Free Graphs}\label{2P2free}

A graph $G=(V,E)$ is a {\em split graph} if $V$ can be partitioned into a clique and an independent set, say $V=C \cup I$ for a clique $C$ and independent set $I$ with $C \cap I =\emptyset$.
In \cite{ChaLiu1993}, the ED problem was solved in linear time for split graphs.

Since a graph is a split graph if and only if it is $\left\{ 2P_2,C_4,C_5 \right\}$-free \cite{FoeHam1977}, $2P_2$-free graphs generalize split graphs.
\begin{theorem}\label{2P2freelin}
The ED problem can be robustly solved in linear time $O(n + m)$ for $2P_2$-free graphs.
\end{theorem}

For showing Theorem \ref{2P2freelin}, we need some definitions and preparing steps.
A set $H$ of at least two vertices of a graph $G$ is called \emph{homogeneous} if $H \not= V(G)$ and every vertex outside $H$ is either adjacent to all vertices in $H$, or to no vertex in $H$.
Obviously, $H$ is homogeneous in $G$ if and only if $H$ is homogeneous in the complement graph $\overline{G}$.
A graph is {\em prime} if it contains no homogeneous set.
A homogeneous set $H$ is \emph{maximal} if no other homogeneous set properly contains $H$.
It is well known that in a connected graph $G$ with connected complement $\overline{G}$, the maximal homogeneous sets are pairwise disjoint and can be determined in linear time (see, e.g., \cite{McCSpi1999}).
The {\em characteristic graph} $G^*$ of $G$ results from $G$ by contracting each of the maximal homogeneous sets $H$ of $G$ to a single representative vertex $h \in H$, and connecting two such vertices by an edge if and only if they are adjacent in $G$.
It is well known that $G^*$ is a prime graph.

Suppose that $G$ is a connected graph having an e.d.\ $D$.
If $\overline{G}$ is not connected, then if $G$ has an e.d.\ $D$, $|D|=1$.
Thus, in this case, we have to test whether $G$ has a universal vertex.
Hence, from now on assume that $G$ and $\overline{G}$ are connected.
Then the characteristic graph $G^*$ is well-defined and prime.

If $G$ admits an e.d.\ $D$ then
\begin{equation}
\text{for every homogeneous set $H$ of $G$: } |H \cap D| \le 1.
\label{eqn:dinhomo}
\end{equation}
\begin{myproof}
Assume that there is a homogeneous set $H$ of $G$ and $d,d' \in D$ with $d\not = d'$ and $d,d' \in H$.
Since $G$ is connected, there is a vertex $x \in V\setminus H$ with $dx \in E$ and $d'x \in E$ -- a contradiction to the e.d.\ property.
\end{myproof}

From now on assume that $G$ is $2P_2$-free. If $G$ has an e.d.\ $D$ then
\begin{equation}
\text{no } d \in D \text{ is in a homogeneous set of } G.
\label{eqn:dmod}
\end{equation}
\begin{myproof}
Assume that there is $d \in D$ in a homogeneous set $H$. Let $x \in H$ be another vertex in $H$.
If $dx \not \in E$, there must be $d' \in D$ with $xd' \in E$.
By (\ref{eqn:dinhomo}), $d' \not \in H$.
Since $H$ is a homogeneous set, $dd' \in E$ -- a contradiction.
Hence, $dx \in E$.

Since $\overline{G}$ is connected, $G$ has no universal vertex and thus $|D|>1$.
Let $d' \in D$.
By (\ref{eqn:dinhomo}), $d' \not\in H$.
Since $G$ is connected, $d'$ has at least one neighbor, say $x'$.
Since $D$ is an e.d., $dd' \not \in E$ and hence $x' \not \in H$.
By the e.d.\ property and since $H$ is a homogeneous set, $xx' \not \in E$.
Hence, $d,x,d',x'$ induce a $2P_2$ in $G$ -- a contradiction.
\end{myproof}
Next we claim:
\begin{equation}
\text{for every $d \in D$ with $|N(d)| \ge 2$, $N(d)$ is a homogeneous set in $G$.}
\label{eqn:dneighbors}
\end{equation}
\begin{myproof}
Assume that $d \in D$ has neighbors $x,y$, and $N(d)$ is not a homogeneous set in $G$.
Then there is a vertex $z \notin N(d)$ distinguishing $x$ and $y$, say $xz \in E$ and $yz \notin E$.
Since $z \notin N(d)$ and, by the e.d.\ property, $z \notin D$, there is a vertex $d' \in D$ with $d' \neq d$ and $d'z \in E$, but now, $d,y,d',z$ induce a $2P_2$, a contradiction.
\end{myproof}

Furthermore,
\begin{equation}
\text{if $D$ is an e.d.\ of $G$, then $D$ is an e.d.\ of $G^*$.}
\label{eqn:GtoGstar}
\end{equation}
\begin{myproof}
Let $D$ be an e.d.\ of $G$.
By \eqref{eqn:dmod}, no $d \in D$ is in a homogeneous set of $G$.
Therefore,
all vertices of $D$ are contained in $G^*$. By construction of $G^*$, set $D$ is an e.d.\ in it.
\end{myproof}

Hence, to find an e.d.\ of a $2P_2$-free graph $G$, by (\ref{eqn:dmod}) and (\ref{eqn:GtoGstar}), it suffices to check if $G^*$ admits an e.d.\ $D^*$ such that no vertex of $D^*$ is in a homogeneous set of $G$.
To do so, we need the following notion:

A {\em thin spider} is a split graph $G=(V,E)$ with partition $V=C \cup I$ into a clique $C$ and an independent set $I$ such that every vertex of $C$ has exactly one neighbor in $I$ and vice versa.
We claim:
\begin{equation}
\text{A nontrivial prime $2P_2$-free graph $G$ has an e.d.\ $\Leftrightarrow$ $G$ is a thin spider.}
\label{spider}
\end{equation}
\begin{myproof}
Obviously, in a thin spider the independent set $I$ is an e.d.
Conversely, let $D$ be an e.d.\ of $G$.
By the e.d.\ property, $D$ is an independent set.

We claim that $|N(d)| = 1$ for every $d \in D$:
Since $G$ is connected, $|N(d)| \ge 1$ holds for all $d \in D$.
Assume that $|N(d)|>1$ for some $d \in D$.
Then by (\ref{eqn:dneighbors}), $N(d)$ is a homogeneous set -- a contradiction.

We claim that $G[V\setminus D]$ is a clique:
If there are $x,x' \in V\setminus D$ with $xx' \not\in E$, there are $d,d' \in D$ with $xd \in E$ and $x'd' \in E$.
Then by the e.d.\ property $d,x,d',x'$ induce a $2P_2$ in $G$ -- a contradiction.

Since every vertex of $V\setminus D$ has exactly one neighbor in $D$, $G$ is a thin spider.
\end{myproof}

Thus, an algorithm for solving the WED problem on $2P_2$-free graphs does the following:
For a given nontrivial connected graph $G$:
\begin{enumerate}[(a)]

\item
Check whether $\overline{G}$ is connected.
If not, then check whether $G$ has a universal vertex.
If not, then $G$ has no e.d.
Otherwise, minimize $\omega(u)$ over all universal vertices $u$ of $G$.

\item ({\em Now $G$ and $\overline{G}$ are connected}.)
Construct the characteristic graph $G^*$ of $G$ and check whether $G^*$ is a thin spider.
If not, then $G$ has no e.d.\ or is not $2P_2$-free.
If $G^*$ is a thin spider, then let $V(G^*)=C \cup I$ be its split partition.
Check if any vertex of $I$ is in a homogeneous set of $G$.
If so, then $G$ has no e.d.\ or is not $2P_2$-free.
Otherwise, $D = I$ is the minimum weight e.d.\ for $G$.
\end{enumerate}

Since modular decomposition can be computed in linear time \cite{McCSpi1999}, Theorem~\ref{2P2freelin} follows.

\section{The WED Problem for $P_5$-Free Graphs}\label{P5freeED}

Since the ED problem is \NP-complete for $P_7$-free graphs, it is interesting to study the complexity of the WED problem for subclasses of $P_7$-free graphs.
We start with $P_5$-free graphs.
Note that for the closely related MWIS problem, its complexity on $P_5$-free graphs is one of the main open problems regarding the complexity of the MWIS problem in hereditary graph classes \cite{LozMos2009,RanSch2010}.

\subsection{A Direct Solution for the WED Problem on $P_5$-Free Graphs}\label{P5freeEDdirect}

\begin{theorem}
The WED problem is solvable in time $O(n m)$ on $P_5$-free graphs in a robust way.
\label{thm:edonp5}
\end{theorem}

To prove Theorem \ref{thm:edonp5}, we need some preparations:
Assume that $G$ admits an e.d.\ $D$.
Let $v\in D$ and let $N_1, N_2,\ldots$ be its distance levels.
If $G$ is $P_5$-free, clearly $N_i=\emptyset$ for all $i>3$.
Moreover, clearly
\begin{equation}
N_1 \cap D = N_2 \cap D = \emptyset.
\label{eqn:n12empty}
\end{equation}
Furthermore,
\begin{equation}
\text{for every edge } yz \in E(G[N_3]): N(y) \cap N_2 = N(z) \cap N_2.
\label{eqn:n2neighbors}
\end{equation}
\begin{myproof}
Assume without loss of generality that there is $x \in \left( N(y) \setminus N(z) \right) \cap N_2$.
Let $w \in N(x) \cap N_1$.
Then $v,w,x,y,z$ is a $P_5$ in $G$---a contradiction.
\end{myproof}
Let $H$ be a component of $G[N_3]$.
By (\ref{eqn:n12empty}), all vertices of $N_3$ must be dominated by vertices in $D\cap N_3$ and by (\ref{eqn:n2neighbors}), all vertices of $H$ have at least one common neighbor in $N_2$.
Hence,
\begin{equation}
H \text{ contains a nonempty set of universal vertices $U$},
\label{eqn:universal}
\end{equation}
and since the choice of a universal vertex of $H$ for $D$ is independent from the choice in the other components of $G[N_3]$,
we may assume that
\begin{equation}
D \text{ contains one vertex of $U$ with minimum weight}.
\label{eqn:universalmin}
\end{equation}
By (\ref{eqn:n12empty}), the vertices of $N_2$ must be dominated by vertices of $N_3$, hence every vertex of $N_2$ has at least one neighbor in $N_3$.
Together with (\ref{eqn:n2neighbors}) and (\ref{eqn:universalmin}) this implies that
\begin{equation}
\text{for all } w \in N_2, N(w) \cap N_3 \text{ is a component of } G[N_3],
\label{eqn:conncomp}
\end{equation}
because otherwise a vertex of $N_2$ would have two neighbors in $D$.

Conversely:
\begin{clai}\label{lma:edonp5correctness}
Suppose that $D$ is a subset of $G$ such that $v\in D$,
for every $w\in N_2$, $N(w) \cap N_3$ is a connected component of $G[N_3]$, and
$D$ contains a universal vertex $u$ of every component $H$ of $G[N_3]$,
then $D$ is an e.d.\ of $G$. \end{clai}

\begin{myproof}
Clearly, the assumptions imply that  $D$ is an independent set.
Moreover, $D$ contains no vertices with common neighbors, because $v$ has distance $3$ to all other vertices of $D$ and if there are two vertices in $D\cap N_3$ with a common neighbor $w$, then $w\in N_2$ by construction,
contradicting the assumption that $N(w) \cap N_3$ is a connected component of $G[N_3]$.
All vertices of $N_1$ are connected to $v$, all vertices in $N_2$ have a neighbor in $D\cap N_3$ and all vertices in $N_3\setminus D$ have a neighbor in $D$. Hence, $D$ is dominating, and thus an e.d.
\end{myproof}

Claim \ref{lma:edonp5correctness} enables us to give the following:

\medskip
\medskip
\hrule
\medskip
\noindent
\textbf{Procedure:} \texttt{Robust-$P_5$-Free-Best-Candidate-for-Vertex}
\begin{enumerate}[(a)]
\item If $N_4 \neq \emptyset$ then \texttt{Stop}---$G$ is not $P_5$-free.
\item Find the components $H_1,\dots,H_k$ of $G[N_3]$, and for every $H_i$ let $U_i$ be the set of universal vertices of $H_i$.
\item Check for every $w \in N_2$ and every $H_i$ if $w$ sees either every or no vertex of $H_i$.
If not then \texttt{Stop}---$G$ is not $P_5$-free.
\item Check for every $w \in N_2$ if there is an $H_i$ such $w$ sees exactly the vertices of $H_i$ in $N_3$.
If not, then $v$ is an unsuccessful choice--- \texttt{Stop and Return $\emptyset$}.
\item Check if every $U_i$ is nonempty.
If not, then $v$ is an unsuccessful choice---\texttt{Stop and Return $\emptyset$}.
\item Let $u_i \in U_i$ of minimum weight for every $U_i$. Set $D=\left\{ v,u_1,\dots,u_k \right\}$.
\texttt{Stop and Return $D$}.
\end{enumerate}
\hrule

\medskip
\begin{lemma}
Algorithm \texttt{Robust-$P_5$-Free-WED} is correct and runs in time $O(n m)$.
\label{lma:edonp5algo}
\end{lemma}
\begin{myproof}
Clearly, Step (a) is correct.
By (\ref{eqn:n2neighbors}) and (\ref{eqn:conncomp}), steps (c) and (d) are correct.
Step (e) is correct by (\ref{eqn:universalmin}), because if there is no universal vertex in some
component of $G[N_3]$, there is no e.d.\ containing $v$.
Hence, the algorithm is correct.

The components of $G[N_3]$ can be computed in linear time using breadth-first-search using Tarjan's algorithm.
When a component is found, its universal vertices can easily be determined by counting their neighbors in the component.
Hence, steps (b) can be done in time $O(n+m)$.
The steps (c) and (d) can be done in linear time in the following way:
Iterate over all vertices of $N_3$ and label its neighbors in $N_2$ with the component of $G[N_3]$ the current vertex is in.
This takes at most $O(m)$ time.
After that, for every vertex of $N_2$ count its labels for the same component of $G[N_3]$ and compare it with the size of the component. If it differs, the check in step (c) fails.
Again, this takes at most $O(m)$ time. Then check if any vertex of $N_2$ is labeled with two or more components.
If so, the check in (d) fails.
This can be done in $O(|N_2|)$ time.
Clearly, steps (e) and (f) can be done in linear time.

This gives an overall runtime of $O(n (n+m))$ which equals $O(n m)$ on connected graphs.
\end{myproof}
This completes the proof of Theorem \ref{thm:edonp5}.

\subsection{Reducing the ED Problem on $P_5$-Free Graphs to the MWIS Problem on Squares}\label{P5freeEDMWIS}

\begin{proposition}\label{midpointP4}
In a $P_5$-free graph $G$, midpoints of an induced $P_4$ are not in any e.d.\ of $G$.
\end{proposition}
\begin{myproof}
Let $G$ be a $P_5$-free graph having an e.d.\ $D$, and let $(a,b,c,d)$ induce a $P_4$ in $G$ with midpoints $b$ and $c$ and endpoints $a,d$. Assume to the contrary that $b \in D$. Then, since $d \notin D$, there is some $d' \in D$ with $dd' \in E$.
Now, by the e.d.\ property, $a,b,c,d,d'$ induce a $P_5$, a contradiction.
\end{myproof}

\begin{theorem}\label{EDP5freesquare}
If graph $G$ is $P_5$-free and has an e.d.\ then $G^2$ is $P_4$-free.
\end{theorem}
\begin{myproof}
Let $G=(V,E)$ be a $P_5$-free graph having an e.d.\ $D$, and assume to the contrary that $G^2$ contains an induced $P_4$ $(a,b,c,d)$.
Then $\delta_G(a,b) \le 2$, $\delta_G(b,c) \le 2$, and $\delta_G(c,d) \le 2$ while $\delta_G(a,c) \ge 3$, $\delta_G(a,d) \ge 3$, and $\delta_G(b,d) \ge 3$.
Since $(a,b,c,d)$ is a $P_4$ in $G^2$, $\delta_G(a,b)=\delta_G(b,c)=\delta_G(c,d)=1$ is impossible. Thus, there are additional vertices of $G$ in the subgraph $G[P]$ which leads to the $P_4$ $(a,b,c,d)$ in $G^2$. If there is only one additional vertex $x \in G$ being adjacent to $b$ and $c$ then $P = (a,b,x,c,d)$ is an induced $P_5$ in $G$, a contradiction. Thus, there are at least two additional vertices $x,y$. If there are only two, say $x,y$, such that $x$ sees $a$ and $b$ and $y$ sees $b$ and $c$ then, since $G$ is $P_5$-free, $xy \in E$ but now $a,x,y,c,d$ induce a $P_5$, a contradiction. Thus, the only remaining cases are the following two:
\begin{enumerate}
\item[(1)] There are two vertices $x,y \in G$ such that $P = (a,x,b,c,y,d)$ is a path in $G$ with $xy \in E$.

\item[(2)] There are three vertices $x,y,z \in G$ such that $P = (a,x,b,y,c,z,d)$ is a path in $G$ with $xy,xz,yz \in E$.
\end{enumerate}

\noindent
{\bf Case (1):} We first claim that none of the vertices $a,x,b,c,y,d$ are in $D$: By Proposition \ref{midpointP4}, $b,c,x,y \notin D$. Then there is $c' \in D$ with $cc' \in E$. Suppose that $a \in D$. Then $c'x \notin E$ by the e.d.\ property. Since $a,x,b,c,c'$ do not induce a $P_5$, $c'b \in E$ follows. Since by Proposition \ref{midpointP4}, $c'$ is not a midpoint of a $P_4$ $d,c',b,x$, it follows that $c'd \notin E$. Since $c',b,x,y,d$ do not induce a $P_5$, $c'y \in E$ follows but now $c'$ is midpoint of $P_4$ $b,c',y,d$, a contradiction. Thus, $a \notin D$ and by symmetry, also $d \notin D$.

\medskip

Now $a,x,b,c,y,d \notin D$. Thus, there is $a' \in D$ with $aa' \in E$. By the distances in $G^2$, $a'$ misses $c$ and $d$, and thus, there is $c' \in D$ with $c'c \in E$ and $c' \neq a'$.
Since $a' \in D$ is not a midpoint of a $P_4$, $a,a',b,c$ do not induce a $P_4$ and thus $a'b \notin E$.
Since $a',a,x,b,c$ do not induce a $P_5$, $a'x \in E$ and thus by the e.d.\ property, $c'x \notin E$.
Since $a' \in D$ is not a midpoint of a $P_4$, $a,a',y,c$ do not induce a  $P_4$ and thus $a'y \notin E$.
Since $a',x,y,c,c'$ do not induce a $P_5$, $c'y \in E$ holds.
Since $a',x,b,c,c'$ do not induce a  $P_5$, $c'b \in E$ but now $b,c',y,d$ is a $P_4$ with midpoint $c'$, a contradiction.

\medskip

\noindent
{\bf Case (2):} Again, we first claim that none of the vertices $a,x,b,y,c,z,d$ are in $D$: By Proposition \ref{midpointP4}, $x,y,z \notin D$.

Suppose that $a \in D$. Then $b,x,y,z \notin D$, and there is a vertex $b' \in D$ with $bb' \in E$. By the distances in $G^2$ and since $a,b,c,d$ is a $P_4$ in $G^2$, $b'd \notin E$, and by the e.d.\ property, $b'x \notin E$ holds. Since $b',b,x,z,d$ is no $P_5$, $b'z \in E$ but now, $b,b'z,d$ is a $P_4$ with midpoint $b'$, a contradiction. Thus, $a \notin D$ and, by symmetry, also $d \notin D$.

Suppose that $b \in D$. Then $a,c,x,y,z \notin D$, and there is a vertex $c' \in D$ with $cc' \in E$. By the distances in $G^2$ and since $a,b,c,d$ is a $P_4$ in $G^2$, $c'a \notin E$, and by the e.d.\ property, $c'x \notin E$ and $c'y \notin E$ holds but now $c',c,y,x,a$ is a $P_5$, a contradiction. Thus, $b \notin D$ and, by symmetry, also $c \notin D$.

\medskip

Now there is $a' \in D$ with $aa' \in E$. Then by the distances in $G^2$, $a'c \notin E$ and $a'd \notin E$, and since $c,y,a',a$ do not induce a $P_4$ with midpoint $a'$, we have $a'y \notin E$. Since $a',a,x,y,c$ do not induce a $P_5$, $a'x \in E$ follows.

Since $c \notin D$ and $a'c \notin E$, there is $c' \in D$ with $cc' \in E$. By the e.d.\ property, $c'a \notin E$ and $c'x \notin E$. Since $a',x,y,c,c'$ do not induce a $P_5$, it follows that $c'y \in E$. Since $d,c',y,x$ do not induce a $P_4$ with midpoint $c'$, $c'd \notin E$ follows. Thus, there is $d' \in D$ with $dd' \in E$ and $d' \neq a'$, $d' \neq c'$. By the e.d.\ property, $d'$ misses $a,x,c,y$. Since $d',d,z,c,c'$ do not induce a $P_5$, $d'z \in E$ or $c'z \in E$ follows. If $c'z \in E$ then $d'z \notin E$ and $a'z \notin E$, and now $d',d,z,x,a'$ induce a $P_5$, and if $d'z \in E$ then $c'z \notin E$ and $a'z \notin E$, and now $a',x,z,c,c'$ induce a $P_5$, a contradiction.
\end{myproof}

Let $T(n,m)$ be the best time bound for constructing $G^2$ from given graph $G$.
 Using the fact that the MWIS and recognition problems are solvable in linear time for $P_4$-free graphs~\cite{CorPerSte1985,CouMakRot2000}, we have, by Lemma \ref{mainequived}:

\begin{corollary}\label{EDP5freeMWISsquare}
For a given $P_5$-free graph $G$, the WED problem can be solved in time $T+O(|E(G^2)|)$.
\end{corollary}

Since $G^2$ can be computed from $G$ using matrix multiplication, this time bound
is incomparable with the $O(n m)$ bound obtained in Theorem \ref{thm:edonp5}.

\medskip

We leave the existence of a linear time algorithm for the (W)ED problem on $P_5$-free graphs as an open problem.


\section{The WED Problem for $\{P_6,S_{1,2,2}\}$-Free Graphs}\label{P6S122free}

Recall that the ED problem is \NP-complete for $P_7$-free graphs, and its complexity is open for $P_6$-free graphs. Let $S_{1,2,2}$ (sometimes called $E$) denote the graph with six vertices, say $a,b,c,d,e,f$, such that $a,b,c,d,e$ induce a $P_5$ with edges $ab,bc,cd,de$ and $f$ is only adjacent to $c$.
Note that the ED problem is \NP-complete for $S_{1,2,2}$-free graphs since it is already \NP-complete for line graphs (and thus for claw-free graphs) as mentioned in Corollary \ref{EDlinegrNPc}.
In this section, as a generalization of the $P_5$-free case, we are going to show:

\begin{theorem}\label{EDP6Efree}
For $\{P_6,S_{1,2,2}\}$-free graphs, the WED problem can be solved in time $O(n^2 m)$ in a robust way.
\end{theorem}

The proof of Theorem \ref{EDP6Efree} needs some preparing steps. Let $G=(V,E)$ be a connected $P_6$-free graph having an e.d.\ $D$.
Let $v \in D$ and consider the distance levels of $G$ with respect to $v$. If $G$ is $P_6$-free then clearly, we have:
\begin{equation}\label{N5empty}
N_5 = \emptyset.
\end{equation}

%
Since $v \in D$, we obviously have:
\begin{equation}\label{N2N3capD=empty}
(N_1 \cup N_2) \cap D = \emptyset.
\end{equation}

Thus, since $D$ is an e.d., no vertex in $N_2$ can be in $D$, but on the other hand, all vertices in $N_2$ have to be dominated; this can be done only by vertices in $N_3$.
We claim:
\begin{equation}\label{DcapN4empty}
D \cap N_4 = \emptyset.
\end{equation}

\noindent
\begin{myproof} 
Assume to the contrary that there is a vertex $w \in D \cap N_4$. Let $c \in N_3$ be a neighbor of $w$, let $b \in N_2$ be a neighbor of $c$ and let $a \in N_1$ be a neighbor of $b$. Then $b$ has to be dominated by a $D$-vertex $d \in N_3$, and since $D$ is an e.d., $cd \notin E$ and $dw \notin E$ but now $v,a,b,c,d,w$ induce an $S_{1,2,2}$, a contradiction.
\end{myproof}

\medskip

We claim:
\begin{equation}\label{atmostoneDN3N4}
\mbox{At most one vertex in } D \cap N_3 \mbox{ has neighbors in } N_4.
\end{equation}

\noindent
\begin{myproof} 
Assume that there are two vertices $d_1,d_2 \in N_3 \cap D$ with neighbors in $N_4$, say $x_i \in N_4$ with $d_ix_i \in E$ for $i=1,2$.
Let $b_i \in N_2$ with $b_id_i \in E$ for $i=1,2$ and let $a_1 \in N_1$ with $a_1b_1 \in E$.
Since $D$ is an e.d., $b_1 \neq b_2$ and $x_1 \neq x_2$ and $d_1$ misses $b_2,x_2$ while $d_2$ misses $b_1,x_1$.
If $x_1x_2 \in E$, $va_1b_1d_1x_1x_2$ is a $P_6$ in $G$, hence, $x_1x_2 \notin E$ holds.
Now if $b_1b_2 \in E$ then there is a $P_6$ in $G$, and if $b_1b_2 \notin E$, there is a $P_6$ as well - a contradiction which shows (\ref{atmostoneDN3N4}).
\end{myproof}

\medskip

Let $R_1,\ldots R_k$ denote the connected components of $G[N_4]$.
Then (\ref{DcapN4empty}) and (\ref{atmostoneDN3N4}) imply that in order to dominate $N_4$, one needs a vertex in $N_3$ which is universal for $N_4$:
\begin{equation}\label{N4dombycommonn}
N_4 \mbox{ can be dominated by some } d \in D \Leftrightarrow \exists x \in N_3 \text { with } N_4 \subseteq N(x).
\end{equation}

Since $G$ is $S_{1,2,2}$-free, we obtain:
\begin{equation}\label{neighbinclus}
\mbox{ If } x \in N_2 \mbox{ is dominated by } d_x \in N_3 \cap D \mbox{ then } N(x) \cap N_3 \subseteq N(d_x) \cap N_3.
\end{equation}

\noindent
\begin{myproof}
Assume that $x \in N_2$ is dominated by $d_x \in N_3 \cap D$ and sees a vertex $r \in N_3\setminus\{d_x\}$ which misses $d_x$; then by the e.d.\ property, $r \notin D$. Let $a \in N_1$ be a common neighbor of $v$ and $x$. Since $rd_x \notin E$ but $r$ has to be dominated by some vertex $d_r \in D\cap N_3$,
it follows by the e.d.\ property that $d_rx \notin E$ but now, $v,a,x,r,d_x,d_r$ induce an $S_{1,2,2}$, a contradiction.
\end{myproof}

\medskip

Consequently, if $x \in N_2$ sees $d_x \in D \cap N_3$, and $y \in N_2$ sees $d_y \in D \cap N_3$ for $d_x \neq d_y$, we obtain:
\begin{equation}\label{nonemptycapinN3}
N(x) \cap N(y) \cap N_3 = \emptyset.
\end{equation}

\noindent
\begin{myproof}
Assume that $x \in N_2$ ($y \in N_2$, respectively) is dominated by $d_x \in D \cap N_3$ ($d_y \in D \cap N_3$, respectively), and $N(x) \cap N(y) \cap N_3 \neq \emptyset$; let $z \in N(x) \cap N(y) \cap N_3$. Then, by (\ref{neighbinclus}), $d_x$ sees $z$ and $d_y$ sees $z$ -- a contradiction to the e.d.\ property.
\end{myproof}

\medskip

This means that vertices $x,y \in N_2$ with a common neighbor in $N_3$ have to be dominated by the same vertex from $D$.

\medskip
Now assume that for $x,y \in N_2$, $N(x) \cap N(y) \cap N_3 = \emptyset$, and let $x$ ($y$, respectively) be dominated by $d_x \in D \cap N_3$ ($d_y \in D \cap N_3$, respectively). We claim:
\begin{equation}\label{emptycapinN3}
d_x,d_y \mbox{ belong to different connected components of } G[N_3].
\end{equation}

\noindent
\begin{myproof}
Assume to the contrary that $d_x$ and $d_y$ are in the same component $Q$ of $G[N_3]$. Then, by the e.d.\ property, the distance between $d_x$ and $d_y$ is at least 3. Let $P=(d_x,u_1,\ldots,u_k,d_y)$ be a shortest path in $Q$ connecting $d_x$ and $d_y$ with $k \ge 2$. Note that $xd_y \notin E$. Let $a \in N_1$ be a common neighbor of $v$ and $x$. If $xu_k \in E$ then $v,a,x,d_x,u_k,d_y$ induce an $S_{1,2,2}$, and if $xu_k \notin E$ then after the last neighbor of $x$ on $P$, there are at least two non-neighbors of $x$, and thus, $v,a,x$ and some vertices of $P$ induce a $P_6$ -- a contradiction.
\end{myproof}

\medskip

We claim:
\begin{equation}\label{onlyoneDinN3}
\mbox{No component in } G[N_3] \mbox{ contains two vertices of } D.
\end{equation}

\noindent
\begin{myproof}
Assume to the contrary that there is a component $Q$ in $G[N_3]$ which contains $d_1,d_2 \in D$, $d_1 \neq d_2$. Both vertices $d_1,d_2$ have neighbors in $N_2$, say
$xd_1 \in E$ with $x \in N_2$ and $yd_2 \in E$ with $y \in N_2$. Then, by the e.d.\ property, $x \neq y$. Then by (\ref{nonemptycapinN3}), $N(x) \cap N(y) \cap N_3 = \emptyset$, and by (\ref{emptycapinN3}), $d_1$ and $d_2$ belong to different components in $N_3$, a contradiction.
\end{myproof}

\medskip

By (\ref{onlyoneDinN3}), components $Q$ in $G[N_3]$ can only be dominated  by universal vertices of $Q$. In other words:
\begin{equation}\label{incomponN3univexist}
\mbox{Every component of } N_3 \mbox{ contains a universal vertex}.
\end{equation}

The problem is how to identify those universal vertices which have to belong to $D$. For this, the following fact is helpful:
Let $u_1,u_2$ be two universal vertices in a component $Q$ of $G[N_3]$. We claim that if their neighborhoods in $N_2$ are incomparable then none of them is in $D$:
\begin{equation}\label{universalincompN3}
\mbox{If } N(u_1) \cap N_2 \nsubseteq N(u_2) \cap N_2 \mbox{ and } N(u_2) \cap N_2 \nsubseteq N(u_1) \cap N_2 \mbox{ then } u_1,u_2 \notin D.
\end{equation}

\noindent
\begin{myproof}
Assume to the contrary that for universal vertices $u_1,u_2$ in a component $Q$ in $G[N_3]$ with incomparable neighborhood in $N_2$, one of them, say $u_1$ is in $D$.
Let $x \in N_2$ see $u_1$ and miss $u_2$, and let $y \in N_2$ see $u_2$ and miss $u_1$.
Then there is $d \in D$ seeing $y$; by (\ref{onlyoneDinN3}), $d$ is in a different component $Q'$ of $G[N_3]$. Let $a \in N_1$ be a common neighbor of $v$ and $x$.
Since $v,a,y,u_2,u_1,d$ do not induce an $S_{1,2,2}$, $ay \notin E$ holds. Since $v,a,x,u_1,u_2,y$ do not induce a $P_6$, $xy \in E$ holds but now, $v,a,x,u_1,y,d$ induce an $S_{1,2,2}$, a contradiction which shows (\ref{universalincompN3}).
\end{myproof}

\medskip

Now let $u_1$ be universal in a component $Q$ of $G[N_3]$, and let $u_2$ be universal in a component $Q'$ of $G[N_3]$, $Q \neq Q'$.
We claim:
\begin{equation}\label{universalindiffcompN3}
N(u_1) \cap N(u_2) \cap N_2 = \emptyset.
\end{equation}

\noindent
\begin{myproof}
Assume to the contrary that there are universal vertices $u_1 \in Q$, $u_2 \in Q'$ with common neighbor $x \in N_2$. If $u_1 \in D$ then, by the e.d.\ property,
$u_2 \notin D$ and thus, there is $d \in D \cap Q'$ with $u_2d \in E$ and $dx \notin E$.  Let $a \in N_1$ be a common neighbor of $v$ and $x$. Now $v,a,x,u_1,u_2,d$ induce $S_{1,2,2}$. Thus, by symmetry, $u_1,u_2 \notin D$ but now, there are vertices $d,d' \in D$ such that $d \in Q$ and $d' \in Q'$, $d \neq u_1$, $d' \neq u_2$. If neither $d$ nor $d'$ sees $x$ then $a,x,u_1,d,u_2,d'$ induce $S_{1,2,2}$, and if one of the $D$ vertices sees $x$, say $dx \in E$ then apply the previous argument by replacing $u_1$ by $d$.
This leads to a contradiction in all cases showing (\ref{universalindiffcompN3}).
\end{myproof}

\medskip

The above conditions lead to the following:

\vbox{
\medskip
\medskip
\hrule
\medskip
\noindent
\textbf{Procedure:} \texttt{Robust-$\{P_6,S_{1,2,2}\}$-Free-Best-Candidate-for-Vertex}
\begin{enumerate}
\item[(a)] If $N_5 \neq \emptyset$ then \texttt{Stop}---$G$ is not $P_6$-free. Otherwise initialize $D:=\{v\}$.

\item[(b)] ({\em Now only $N_1,N_2,N_3,N_4$ can be nonempty.})
\begin{enumerate}

\item[(b.1)] Determine the connected components $R_1,\ldots,R_k$ of $G[N_4]$.

\item[(b.2)] If $k>0$, determine the set $M$ of vertices in $N_3$ that are universal for $N_4$.
If $M=\emptyset$ then $v$ is an unsuccessful choice--- \texttt{Stop and Return $\emptyset$}.
\end{enumerate}

\item[(c)] ({\em Now there is a universal vertex for $N_4$ in $N_3$.})
\begin{enumerate}

\item[(c.1)] Determine the connected components $Q_1,\ldots,Q_{\ell}$ of $G[N_3]$.

\item[(c.2)] For all $i \in \{1,\ldots,\ell\}$, determine the set $U_i$ of universal vertices in $G[Q_i]$. If there is $i \in \{1,\ldots,\ell\}$ with $U_i=\emptyset$
then $v$ is an unsuccessful choice---\texttt{Stop and Return $\emptyset$}.

\item[(c.3)]
For all $i \in \{1,\ldots,\ell\}$, let $u^M_i$ be a vertex of $U_i \cap M$ of minimum weight and let $u_i$ be a vertex of $U_i \setminus M$ of minimum weight.
Set $D_i := D \cup \left\{ u^M_i \right\} \cup \left( \left\{ u_1,\dots,u_\ell \right\} \setminus \left\{ u_i \right\} \right)$.
If $D_i$ does not exist for all $i \in \left\{ 1,\dots,\ell \right\}$, because one of the appropriate vertices to choose does not exist, $v$ is an unsuccessful choice---\texttt{Stop and Return $\emptyset$}.
Otherwise, \texttt{Stop and Return $D_i$} for a $D_i$ of minimum weight.
\end{enumerate}
\end{enumerate}
\hrule
}

\medskip
\begin{lemma}\label{EDP6Efreelemma}
Algorithm \texttt{Robust-$\{P_6,S_{1,2,2}\}$-Free-WED} is correct and runs in time $O(n^2m)$.
\end{lemma}

\noindent
\begin{myproof}
{\em Correctness}:
By (\ref{N5empty}), step (a) is correct. Step (b.1) has to construct the connected components in $G[N_4]$.
By (\ref{N4dombycommonn}), step (b.3) is correct. Then, by (\ref{incomponN3univexist}), one needs to determine the connected components of $G[N_3]$ and to find out whether each of them has a universal vertex. By (\ref{universalincompN3}) and (\ref{universalindiffcompN3}), it is correct to choose for every component a universal vertex with maximum neighborhood in $N_2$.

Finally, one has to check once more for every $v$ whether $D_v$ is an e.d.\ of finite weight. If $G$ has one then the algorithm finds the optimal result.

\smallskip
\noindent
{\em Time bound}: The algorithm has to be carried out for every $v \in V$ which is a factor $n$.
For each round, the time bound for (a), (b), and (c) is $O(n+m)$ except (c.3), where we need time $O(nm)$.
Thus, altogether, the time bound is $O(n^2m)$.
\end{myproof}

\noindent
This finally shows Theorem \ref{EDP6Efree}.

\section{The WED Problem for $\{2P_3,S_{1,2,2}\}$-Free Graphs}\label{P3P3S122free}

Recall that the ED problem is \NP-complete for $2P_3$-free as well as for $S_{1,2,2}$-free graphs.
In this section, we give a robust polynomial time algorithm for the WED problem on $\{2P_3,S_{1,2,2}\}$-free graphs.

\begin{theorem}\label{ED2P3S122free}
For $\{2P_3,S_{1,2,2}\}$-free graphs, the WED problem can be solved in time $O(n^5)$ in a robust way.
\end{theorem}

For showing Theorem \ref{ED2P3S122free}, we need some preparing steps.
Assume that $G$ has an e.d.\ $D$.
Let $v \in D$ and $N_1,N_2,\dots$ its distance levels.
If $G$ is $2P_3$-free, we have:
\begin{equation}\label{N6empty}
\mbox{For all } k \ge 6, N_k = \emptyset.
\end{equation}

Let $R:=V \setminus (\{v\} \cup N_1 \cup N_2)$.
We distinguish between the following cases:

\smallskip

\noindent
{\bf Case 1.} $v$ is midpoint of a $P_3$, say $(x,v,y)$.

Since $G$ is $2P_3$-free, $R$ is $P_3$-free, i.e., the disjoint union of some cliques, say $Q_1,\ldots,Q_k$, $k \ge 0$.
If $R=\emptyset$ and $N_2 \neq \emptyset$, then $G$ has no e.d.\ $D$ with $v \in D$.
If $R=N_2 = \emptyset$, then the only e.d.\ $D$ with $v \in D$ is $D =\{v\}$.
Thus let $k \ge 1$. Vertices in $N_2$ can only be dominated by vertices in $R$, and obviously, for every $i$, $1 \le i \le k$:
\begin{equation}\label{DQi}
|D \cap Q_i|=1
\end{equation}

Assume that for $z \in N_2$, $zd \in E$ with $d \in D \cap Q_1$.
Then by (\ref{DQi}) and the e.d.\ property, $z$ has a non-neighbor in every $Q_i$, $i \ge 2$.
If $z$ has a neighbor and a non-neighbor in some $Q_i$, $i \ge 2$, say $zu \in E$ for $u \in Q_2$ and $zw \notin E$ for $w \in Q_2$ then $v,x,z,d,u,w$ induce an $S_{1,2,2}$ for some $x \in N(z) \cap N_1$, a contradiction.
Since $D$ is an e.d., every $z \in N_2$ must see a vertex of some $Q_i$. Thus:
\begin{equation}\label{N2Qi}
\mbox{ If } z \in N_2 \mbox{ sees } Q_i \mbox{ then it misses all } Q_j, j \neq i.
\end{equation}

This implies in Case 1:

\begin{proposition}\label{case12P3S122free}
$G$ has an e.d.\ $D$ with $v \in D$ if and only if for every $i$, $1 \le i \le k$, $Q_i$ contains a vertex $q_i$ with maximum neighborhood among all vertices in $Q_i$ with respect to $N_2$, and $N(q_i) \cap N_2, i \in \{1,\ldots,k\}$, form an exact cover of $N_2$.
\end{proposition}

From now on, in Cases 2 and 3, for all $d \in D$, $d$ is not a midpoint of any $P_3$, i.e., $d$ is simplicial.

\smallskip

\noindent
{\bf Case 2.} $N_4 \neq \emptyset$.

Since now every $d \in D$ is simplicial, we have:
\begin{equation}\label{DN2N4}
\mbox{For all } d \in D \cap N_3, N(d) \cap N_4 = \emptyset.
\end{equation}

Since $G$ is $S_{1,2,2}$-free:
\begin{equation}\label{DcapN4empty2}
D \cap N_4 = \emptyset.
\end{equation}
\begin{myproof}
Assume there is $d \in N_4 \cap D$.
Let $y \in N(d) \cap N_3$ and $x \in N(y) \cap N_2$.
Since $x$ is dominated by $D$, there is $d' \in N_3 \cap N(x) \cap D$.
Since $D$ is efficient, $dd', d'y \not\in E$.
Then $v,w,x,y,d,d'$ induce an $S_{1,2,2}$ in $G$ for some $w \in N(x)\cap N_1$---a contradiction.
\end{myproof}

Thus, vertices in $N_4$ can only be dominated by $D$-vertices in $N_5$, i.e., if $N_4 \neq \emptyset$ then $N_5 \neq \emptyset$.

Let $d' \in D \cap N_5$ and $xd' \in E$ for some $x \in N_4$ and  $xy \in E$ for some $y \in N_3$.
By (\ref{DN2N4}), $y \notin D$. Let $d \in D \cap N_3$ and $bd \in E$ for some $b \in N_2$ and $ab \in E$ for some $a \in N_1$.
Since $v,a,b, y,x,d'$ induce no $2P_3$, $b$ sees $y$.
Since $v,a,b,d,y,x$ induce no $S_{1,2,2}$, $d$ sees $y$.
Now assume that $d$ misses some $b' \in N_2$. Let $a' \in N_1$ be a neighbor of $b'$. Since $v,a',b',y,x,d'$ is no $2P_3$, $b'y \in E$ but now, $a',b',y,x,d,d'$ induce an $S_{1,2,2}$, a contradiction.
This shows
\begin{equation}\label{dunivN2}
d \mbox{ is universal for } N_2.
\end{equation}

This implies
\begin{equation}\label{DcapN3one}
|D \cap  N_3|=1.
\end{equation}

In particular, $N_2 \cup N_3$ can be dominated if and only if there is a vertex in $N_3$ which is universal for $N_2 \cup N_3$. The remaining part $N_4 \cup N_5$ can be treated separately: Since $G$ is $2P_3$-free, we have:
\begin{equation}\label{N4N5cliques}
N_4 \cup N_5 \mbox{ is the disjoint union of some cliques.}
\end{equation}

Thus, in order to obtain an e.d., for every clique $Q$ in $G[N_4 \cup N_5]$, choose a vertex from $Q \cap N_5$ (if possible).

\smallskip

\noindent
{\bf Case 3.} $N_4 = \emptyset$.

Let $D \cap N_3=\{d_1,\ldots,d_k\}$. Since all $D$-vertices are simplicial, $N(d_i)$ is a clique for every $i \in \{1,\ldots,k\}$. However, this does not yet mean that $G[N_3]$ is the disjoint union of cliques since there might be edges between $N(d_i)$ and $N(d_j)$ for $i \neq j$.

Since $G$ is $S_{1,2,2}$-free, every vertex $x \in N_2$ seeing a vertex $d_i \in D$ misses $N[d_j] \cap N_3$ for every $j \neq i$.

Now suppose that $k \ge 3$ and there is an edge $uw \in E$ for $u \in N(d_2) \cap N_3$ and $w \in N(d_3) \cap N_3$. Let $x \in N_2$ with $xd_1 \in E$ and $a \in N_1$ with $ax \in E$. Then $v,a,x,d_2,u,w$ induce $2P_3$, a contradiction. Thus:
\begin{equation}\label{kge3N3cliques}
\mbox{ If } k \ge 3 \mbox{ then } G[N_3] \mbox{ is the disjoint union of cliques.}
\end{equation}

In the other case, $G[N_3]$ must be a co-bipartite subgraph, say with cliques $Q_1$ and $Q_2$, and we can check for every pair of vertices $x \in Q_1,y \in Q_2$ whether $\{v,x,y\}$ is an e.d.\ of $G$.

The above conditions lead to the following:

\medskip
\medskip
\hrule
\medskip
\noindent
\textbf{Procedure:} \texttt{Robust-$\{2P_3,S_{1,2,2}\}$-Free-Best-Candidate-for-Vertex}
\begin{enumerate}
\item[(a)]
If $N_2=\emptyset$, \texttt{Stop and Return $\left\{ v \right\}$}.
If $N_6 \neq \emptyset$ then \texttt{Stop}---$G$ is not $2P_3$-free.

\item[(b)] ({\em Now only $N_1,N_2,N_3,N_4,N_5$ can be nonempty.}) If $v$ is not simplicial then let $R:=V \setminus (\{v\} \cup N_1 \cup N_2)$; if $R=\emptyset$ then
    $v$ is an unsuccessful choice---\texttt{Stop and Return $\emptyset$}.
Otherwise, check whether $R$ is the disjoint union of some cliques, say $Q_1,\ldots,Q_k$, $k \ge 1$.
If not, then \texttt{Stop}---$G$ is not $2P_3$-free.
Otherwise, for every $i \in \{1,\ldots,k\}$, choose a vertex $d_i \in Q_i$ of maximum degree in $N_2$ and minimum weight; let $D:=\{v,d_1,\ldots,d_k\}$.
\texttt{Stop and Return $D$}.

\item[(c)] ({\em Now $v$ and all other vertices in $D$ are simplicial.})

\begin{enumerate}
\item[(c.1)] If $N_4 \neq \emptyset$ then check whether there is a universal vertex in $G[N_2 \cup N_3]$; if not then
then \texttt{Stop}---either $v$ is an unsuccessful choice or $G$ is not $\{2P_3,S_{1,2,2}\}$-free.
Otherwise, choose such a
universal vertex $u$ in $G[N_2 \cup N_3]$ of minimum weight and let $D:=\{v,u\}$. Check whether $G[N_4 \cup N_5]$ is a  disjoint union of cliques. If not then \texttt{Stop}---$G$ is not $2P_3$-free.
Otherwise, let $Q_1,\ldots,Q_k$ be the cliques in $G[N_4 \cup N_5]$. For every $Q_i$, choose a vertex $d_i$ of minimum weight in $Q_i \cap N_5$ if possible---otherwise $v$ is an unsuccessful choice---\texttt{Stop and Return $\emptyset$}.
Let $D:=D \cup \{d_1,\ldots,d_k\}$. \texttt{Stop and Return $D$}.

\item[(c.2)] ({\em Now $N_4 = \emptyset$}.)
\begin{enumerate}
\item[(c.2.1)] Check whether $G[N_3]$ is a disjoint union of cliques, say $Q_1,\ldots,Q_k$.
If yes, proceed analogously to step (b).
\item[(c.2.2)] Check whether $G[N_3]$ is co-bipartite, say with cliques $Q_1$ and $Q_2$ (and possibly edges between $Q_1$ and $Q_2$). If yes, then for every pair of vertices $x \in Q_1$, $y \in Q_2$, check whether $D=\{v,x,y\}$ is an e.d.\ of $G$. If none of them is an e.d.\ then $v$ is an unsuccessful choice---\texttt{Stop and Return $\emptyset$}.
    Otherwise choose one of minimum weight.
\item[(c.2.3)] Finally, if neither (c.2.1) nor (c.2.2) holds true
then \texttt{Stop}---$G$ is not $\{2P_3,S_{1,2,2}\}$-free.
\end{enumerate}
\end{enumerate}
\end{enumerate}
\hrule

\begin{lemma}\label{ED2P3Efreelemma}
Algorithm \texttt{Robust-$\{2P_3,S_{1,2,2}\}$-Free-WED} is correct and runs in time $O(n^5)$.
\end{lemma}
\begin{myproof}
{\em Correctness}:
Clearly, Step (a) is correct.
The correctness of Step (b) follows from (\ref{DQi}) and (\ref{N2Qi}).
Step (c.1) is correct by (\ref{DN2N4})--(\ref{N4N5cliques}).
The correctness of Step (c.2) follows from (\ref{N2Qi}) and the discussion of Case 3.

\smallskip
\noindent
{\em Time bound}:
The algorithm has to be carried out for every $v \in V$ which is a factor $n$.
For each round, the time bound for (a) is $O(n+m)$.
For given $v$, it can be checked in time $O(n+m)$ whether $v$ is simplicial and whether $R$ is the disjoint union of cliques $Q_1,\ldots,Q_k$.
The neighborhood of every clique $Q_i$ in $N_2$ can be determined in time $O(n+m)$, and it can be determined in the same time bound whether $Q_i$ has a vertex with this neighborhood in $N_2$.
Step (c.1) can be done in linear time $O(n+m)$ in a very similar way.
The steps of (c.2) can be done either in time $O(n+m)$ or, in the case of a co-bipartite subgraph in (c.2.2), at most $n^2$ pairs have to be checked which can be done in time $O(n^2(n+m))$.
Thus, altogether, a time bound for the algorithm is $O(n^5)$.
\end{myproof}

This finally shows Theorem \ref{ED2P3S122free}.

\section{The WED Problem for $(P_2+P_4)$-Free Graphs}\label{P2P4free}

We know by Section \ref{2P2free} that the ED problem is linear time solvable on $2P_2$-free graphs and that is is \NP-complete on $2P_3$-free graphs.
Hence, it is interesting to analyze the complexity on graph classes in between.
We do that by showing that ED is polynomial time solvable on $(P_2+P_4)$-free graphs, which implies that is it polynomial time solvable on $(P_2+P_3)$-free graph, a proper superclass of $2P_2$-free graphs and a proper subclass of $2P_3$-free graphs.

\begin{theorem}\label{thm:p2p4freenm}
The WED problem can be solved on $(P_2+P_4)$-free graphs in time $O(n m)$ in a robust way.
\end{theorem}

The proof of Theorem~\ref{thm:p2p4freenm} needs some preparations:
Assume that $G$ admits an e.d.\ $D$.
Let $v \in D$ and let $N_1,N_2,\dots$ be its distance levels.
If $G$ is $(P_2+P_4)$-free, then clearly $N_6 = \emptyset$.

Let $R=N_3 \cup N_4 \cup N_5$.
Then $V=\left\{ v \right\}\cup N_1 \cup N_2 \cup R$ is a partition of $V$.
Clearly $(N_1 \cup N_2) \cap D = \emptyset$, which implies
\begin{equation}
\text{every vertex of $N_2$ has exactly one neighbor in $N_3 \cap D$.}
\label{eqn:n2hasneighbor}
\end{equation}
If $G$ is $(P_2+P_4)$-free, $G[R]$ is a cograph.
For cographs one can easily check:
\begin{clai}
Let $G=(V,E)$ be a cograph.
Then $D \subseteq V$ is an e.d.\ of $G$ if and only if $D$ consists of exactly one universal vertex of every component of $G$.
\label{clai:cograph}
\end{clai}
Since $N_2 \cap D = \emptyset$, $D\setminus\{v\}$ is an e.d.\ of $G[R]$.

Let $Q_i$ be a component of $G[R]$ and $U_i$ the set of universal vertices of $Q_i$.
By Claim \ref{clai:cograph},
\begin{equation}
|D \cap Q_i| = 1 \text{ and } D \cap Q_i \subseteq U_i.
\label{eqn:chooseU}
\end{equation}
If $U_i \cap N_3 = \emptyset$, then for all $y \in U_i$
\begin{equation}
D' = (D \setminus U_i) \cup \left\{ y \right\} \text{ is an efficient dominating set of $G$.}
\label{eqn:independentU}
\end{equation}
\begin{myproof}
By (\ref{eqn:chooseU}), there is $y \in U_i \cap D$.
Let $y' \in U_i$.
Since $U_i \cap N_3 = \emptyset$ and $y$ and $y'$ are universal in $Q_i$, $N[y]=N[y']$.
Hence, replacing $y$ by $y'$ in $D$ leads to an e.d.\ of $G$.
\end{myproof}
%

Now let ${\cal Q} = \{Q_1,\dots,Q_k\}$ be the set of those components of $G[R]$ that have a universal vertex in $N_3$, and let $U_i$ be the set of universal vertices of $Q_i$ that are in $N_3$, for all $i \in \left\{ 1,\dots,k \right\}$.
Note that there is at least one component of this kind, because otherwise, by (\ref{eqn:chooseU}), the vertices of $N_2$ cannot be dominated---a contradiction to the existence of $D$.

This implies that for every component $Q$ of $G[R]$ with $Q \not \in \left\{ Q_1,\dots,Q_k \right\}$, we have $|Q \cap (N_4\cup N_5)|=1$, because otherwise $G$ contains an induced $P_2+P_4$, consisting of a $P_2$ inside $Q$ and a $P_4$ from $v$ along the distance levels to any vertex of $Q_1$ in $N_3$.
Consequently, every such component satisfies $Q \cap N_5=\emptyset$ and $|Q \cap N_4|=1$.


Hence, let $D_4 := N_4 \setminus (Q_1 \cup \dots \cup Q_k)$.
Clearly, $D_4$ efficiently dominates all vertices in components of $G[R]$ different from $Q_1,\dots,Q_k$.

If $k=1$, then
there exists a vertex
$y \in U_1$ with $N_2 \subseteq N(y)$
such that
\begin{subequations}
\begin{equation}
D =  \left\{ v, y \right\} \cup D_4\,,
\label{eqn:k1}
\end{equation}
and for all $y \in U_1$ with $N_2 \subseteq N(y)$,
\begin{equation}
D' =  \left\{ v, y \right\} \cup D_4 \text{ is an efficient dominating set of $G$}\,.
\label{eqn:k1_2}
\end{equation}
\end{subequations}

\begin{myproof}
The fact that $D$ satisfies~\eqref{eqn:k1} follows from the observation that every component $Q$ of $G[R]$ other than $Q_1$ must be dominated by the unique vertex in $Q\cap N_4$, while
all vertices in $N_2$ must be dominated by the same vertex in $U_1$.

Condition~\eqref{eqn:k1_2} follows from Claim \ref{clai:cograph} and $N_2 \cap D' = \emptyset$.
\end{myproof}


From now assume that $k \ge 2$.
Then for all $x \in N_2$ and all $i \in \left\{ 1,\dots,k \right\}$
\begin{equation}
\text{$x$ misses at most one vertex of $U_i$.}
\label{eqn:onemiss}
\end{equation}

\begin{myproof}
Conversely, assume without loss of generality that
there is a vertex $x \in N_2$ that misses two vertices $z,z' \in U_1$.
Let $w \in N_1$ and $y \in U_2$ be two neighbors of $x$.
Then $v,w,x,y$ together with $z,z'$ induce a $P_2+P_4$ in $G$---a contradiction.
\end{myproof}


By (\ref{eqn:onemiss}), if $U_i \cap N_4 \not= \emptyset$ for some $i \in \left\{ 1,\dots,k \right\}$, then $|U_i \cap N_4| = 1$, so let us denote this vertex by $z_i$.
If $U_i \cap N_4 \not= \emptyset$ for all $i \in \left\{ 1,\dots,k \right\}$, then there is $j \in \left\{ 1,\dots,k \right\}$ such that
\begin{subequations}
\begin{equation}
D = (\left\{ v, y_j, z_1, \dots, z_k \right\} \setminus \left\{ z_j \right\}) \cup D_4 \text{ for some } y_j \in U_j \cap N_3\,,
\label{eqn:alln4_1}
\end{equation}
and for every $\ell \in \left\{ 1,\dots,k \right\}$ and every $y_\ell \in U_\ell \cap N_3$,
\begin{equation}
D' := (D \setminus \left\{ y_j, z_\ell \right\}) \cup \left\{ y_\ell, z_j \right\} \text{ is an efficient dominating set of $G$}.
\label{eqn:alln4_2}
\end{equation}
\end{subequations}

\begin{myproof}
If for some $i \in \left\{ 1,\dots,k \right\}$ it holds that $U_i \cap N_4 \not = \emptyset$, then, by (\ref{eqn:onemiss}), $U_i \setminus \left\{ z_i \right\}$ has a join to $N_2$.
Hence, if $(U_i \setminus \left\{ z_i \right\}) \cap D \not= \emptyset$, then, by (\ref{eqn:chooseU}), $z_j \in D$ for all $j \in \left\{ 1,\dots,k \right\}, j \not= i$, because otherwise, $D$ is not efficient.
Since all vertices of $U_i \cap N_3$ have the same neighborhood, every such choice leads to an e.d., if at least one exists.
This proves (\ref{eqn:alln4_1}) and (\ref{eqn:alln4_2}).
\end{myproof}


If there is exactly one $j \in \left\{ 1,\dots,k \right\}$ with $U_j \cap N_4 = \emptyset$, then
\begin{subequations}
\begin{equation}
D = \left\{ v, y_j, z_1, \ldots, z_{j-1},z_{j+1},\ldots, z_k \right\} \cup D_4 \text{ for some } y_j \in U_j \text{ with } N_2 \subseteq N(y_j)\,,
\label{eqn:almostalln4_1}
\end{equation}
and for every $y_j' \in U_j$ with $N_2 \subseteq N(y_j')$,
\begin{equation}
D' := D \setminus \left\{ y_j \right\} \cup \left\{ y_j' \right\} \text{ is an efficient dominating set of $G$}.
\label{eqn:almostalln4_2}
\end{equation}
\end{subequations}
\begin{myproof}
If there is exactly one $j \in \left\{ 1,\dots,k \right\}$ with $U_j \cap N_4 = \emptyset$, then, by (\ref{eqn:chooseU}), $|U_j \cap D| = 1$, say $y \in U_j \cap D$.
Because $D$ is efficient, we must have $z_i \in D$ for all $i \in \left\{ 1,\dots,k \right\}\setminus\{j\}$, since otherwise
any vertex $w$ in $N_2\cap N(y_j)$ would miss at least two vertices in $U_i$ (one in $D\cap U_i$ and $z_i$).
Since all vertices of $U_j$ dominate the vertices of $Q_j$, every choice of a vertex that dominates all vertices of $N_2$ leads to an e.d., if at least one exists.
This proves (\ref{eqn:almostalln4_1}) and (\ref{eqn:almostalln4_2}).
\end{myproof}


From now assume without loss of generality that $U_1 \cap N_4 = \emptyset$ and $U_2 \cap N_4 = \emptyset$.
Assume that there is a vertex $x \in N_2$ such that $U_i \subseteq N(x)$ for some $i \in \left\{ 1,\dots,k \right\}$.
Since then $U_i \cap N_4 = \emptyset$, assume without loss of generality that $i=1$.
Since $D$ is efficient and $|D\cap U_j| = 1$ for all $j\in \{1,\ldots, k\}$ and by (\ref{eqn:onemiss}),
we conclude that for all $j\in \{2,\ldots, k\}$, vertex $x$ misses exactly one vertex in $U_j$.
Let $y_j$ denote the unique vertex missed by $x$ in $U_j$.
Since $U_2 \cap N_4 = \emptyset$, $y_2$ has a neighbor $x' \in N_2$.
Then $x'$ has a unique non-neighbor $y_1 \in U_1$ and
\begin{equation}
D = \left\{ v, y_1, y_2, \dots, y_k \right\} \cup D_4.
\label{eqn:oneseesall}
\end{equation}
\begin{myproof}
By (\ref{eqn:chooseU}) and since $D$ is efficient, $y_2,\dots,y_k \in D$.
Since $y_2 \in D$, $x'$ has a (unique) non-neighbor $y_1 \in U_1$ and $U_1 \cap D = \left\{ y_1 \right\}$, otherwise $D$ is not efficient.
\end{myproof}
%


From now assume that every vertex of $N_2$ misses exactly one vertex of every $U_i$.
Then there exist  vertices $x,x' \in N_2$, $a,b \in U_1$ and $c,d \in U_2$ such that $x$ sees $a$ and $c$ but misses $b$ and $d$, and $x'$ sees $b$ and $d$ and misses $a$ and $c$, and, for all $i \in \left\{ 3,\dots,k \right\}$, $x$ and $x'$ have the same non-neighbor $y_i \in U_i$, and, either
\begin{equation}
\begin{array}{c}
D = \left\{ v, a, d, y_3, \dots, y_k \right\} \cup D_4\\
\text{or}\\
D = \left\{ v, b, c, y_3, \dots, y_k \right\} \cup D_4.
\end{array}
\label{eqn:differentmisses}
\end{equation}
\begin{myproof}
First we show the existence of $x$, $x'$, $a$, $b$, $c$, and $d$.
Since $U_1 \cap N_4 = \emptyset$, there are distinct vertices $w, w' \in N_2$ and $y_1, y_1' \in U_1$ such that $w$ misses $y_1$ and $w'$ misses $y_1'$.
Let $y_2$ be the non-neighbor of $w$ in $U_2$ and let $y_2'$ be the non-neighbor of $w'$ in $U_2$.
If $y_2 \not= y_2'$, set $x:=w$, $x':=w'$, $a:=y_1'$, $b:=y_1$, $c:=y_2'$ and $c:=y_2$ and we are done.
Hence, assume that $y_2=y_2'$.
Since $U_2 \cap N_4 = \emptyset$, $y_2$ has a neighbor $w'' \in N_2$.
Let $y_2''$ be the non-neighbor of $w''$ in $U_2$ and let $y_1''$ be the non-neighbor of $w''$ in $U_1$.
Then either $x:=w$, $x':=w''$, $a:=y_1''$, $b:=y_1$, $c:=y_2''$, $d:=y_2$ or $x:=w'$, $x':=w''$, $a:=y_1''$, $b:=y_1'$, $c:=y_2''$, $d:=y_2$ fulfills the mentioned conditions.

Now assume that $\left\{ a,b \right\} \cap D = \emptyset$.
Then $x$ is dominated by a vertex $y \in U_1$ that also dominates $x'$.
Since every vertex of $U_2$ sees $x$ or $x'$, $D$ is not efficient by (\ref{eqn:chooseU}) -- a contradiction.

Hence, assume that $a \in D$.
Then $d \in D$, because otherwise $D$ would not be efficient.
Analogously, if $b \in D$, then $c \in D$.

Since in both cases $x$ and $x'$ are dominated, by (\ref{eqn:onemiss}) and (\ref{eqn:chooseU}), $x$ and $x'$ have the same non-neighbor $y_i \in U_i$ and $y_i \in D$ for all $i \in \left\{ 3,\dots,k \right\}$.
\end{myproof}

Summarizing, given a $(P_2+P_4)$-free graph $G$ and a vertex $v$, every e.d.\ of $G$ that contains $v$ has the form given in either (\ref{eqn:k1}), (\ref{eqn:alln4_1})---(\ref{eqn:alln4_2}), (\ref{eqn:almostalln4_1})---(\ref{eqn:almostalln4_2}), (\ref{eqn:oneseesall}), or (\ref{eqn:differentmisses}).

\medskip
\medskip
\hrule
\medskip
\noindent
\textbf{Procedure:} \texttt{Robust-$(P_2+P_4)$-Free-Best-Candidate-for-Vertex}
\begin{enumerate}
\item[(a)] Set $D:=\left\{ v \right\}$ and $R:=N_3 \cup N_4 \cup N_5$.

\item[(b)] If $N_2 = \emptyset$, \texttt{Stop and Return $D$}; if $N_6 \not= \emptyset$ or $G[R]$ is not a cograph, then $G$ is not $(P_2+P_4)$-free, \texttt{Stop}.

\item[(c)] Determine the set $\cQ$ of components of $G[R]$.
For every $Q \in \cQ$, determine the set $U$ of universal vertices of $Q$ and check if $U \cap N_3 = \emptyset$.
If so:
\begin{enumerate}
\item[(c.1)] If $|Q \cap N_4| > 1$: If $|\cQ|=1$, then $v$ is an unsuccessful choice---\texttt{Stop and Return $\emptyset$}.
Otherwise, $G$ is not $(P_2+P_4)$-free, \texttt{Stop}.
\item[(c.2)] Set $D := D \cup U$ and $\cQ := \cQ \setminus \left\{ Q \right\}$.
\end{enumerate}

\item[(d)] Let $\cQ = \{Q_1,\dots,Q_k\}$ and let the corresponding sets of universal vertices be $U_1,\dots,U_k$.
If $k=0$, then $v$ is an unsuccessful choice--- \texttt{Stop and Return $\emptyset$}.

\item[(e)] For all $x \in N_2$ and all $i \in \left\{ 1,\dots,k \right\}$ calculate the number $n_i(x)=|N(x)\cap U_i|$; for all $y \in U_1 \cup \dots \cup U_k$, calculate the number $m(y) = |N(y) \cap N_2|$.

\item[(f)] If $k=1$:\\
Let $U_1'$ contain all vertices $y \in U_1$ with $m(y)=|N_2|$.
If $U_1'=\emptyset$, then $v$ is an unsuccessful choice
    ---\texttt{Stop and Return $\emptyset$}.
Otherwise, choose a minimum weight vertex $y \in U_1'$, set $D:=D \cup \left\{ y \right\}$.
\texttt{Stop and Return $D$}.

\item[(g)] Check if there is $x \in N_2$ such that $n_i(x) < |U_i|-1$ for any $i \in \left\{ 1,\dots,k \right\}$. If so, then $G$ is not $(P_2+P_4)$-free, \texttt{Stop}.

\item[(h)] Let $Z$ contain all vertices $z \in U_1 \cup \dots \cup U_k$ with $m(z)=0$ and let $z_i$ denote the vertex in $Z \cap U_i$, if it exists.

\item[(i)] If $|Z| = k$:\\
Choose $i \in \left\{ 1,\dots,k \right\}$ and $y_i \in U_i\setminus\{z_i\}$ so that $\omega(y_i)- \omega(z_i)$ is minimal.
Set $D \cup \left\{ y_i \right\} \cup (Z \setminus \left\{ z_i \right\})$.
\texttt{Stop and Return $D$}.

\item[(j)] If $|Z| = k-1$:\\
Say $U_1 \cap Z = \emptyset$.
For the best choice of $y \in U_1$ such that $m(y)=|N_2|$, set $D:=D \cup \left\{ y \right\} \cup Z$.
\texttt{Stop and Return $D$}.

\item[(k)] If there is a vertex $x \in N_2$ with $n_i(x)=|U_i|$ for some $i \in \left\{ 1,\dots,k \right\}$:\\
Say $i=1$.
For all $i \in \left\{ 2,\dots,k \right\}$, let $y_i$ denote the non-neighbor of $x$ in $U_i$, if it exists, and, for the
minimum weight vertex $y \in U_1$ such that $m(y)=|N_2|$, set $D:=D \cup \left\{ y,y_2,\dots,y_k \right\}$.
\texttt{Stop and Return $D$}.

\item[(l)] Check if there are vertices $x,x' \in N_2$, $a,b \in U_i$ and $c,d \in U_j$ such that $i\neq j$,
$xa \in E$, $xc \in E$, $x'b \in E$, $x'd \in E$ and $xb \not\in E$, $xd \not\in E$, $x'a \not\in E$, $x'c \not\in E$:\\
If such vertices do not exist, then $v$ is an unsuccessful choice---\texttt{Stop and Return $\emptyset$}.
Otherwise, say $i=1$ and $j=2$.
For $r\in \{3,\ldots, k\}$, let $y_r$ be the common non-neighbor of $x$ and $x'$ in $U_r$, if it exists.
If any of these non-neighbors does not exist, then $v$ is an unsuccessful choice---\texttt{Stop and Return $\emptyset$}.
If $\omega(a)+\omega(d) < \omega(b)+\omega(c)$, then set $D:=D \cup \left\{ a,d,y_3,\dots,y_k \right\}$, else set $D:= D \cup \left\{ b,c,y_3\dots,y_k \right\}$. \texttt{Stop and Return $D$}.
\end{enumerate}
\hrule

\begin{lemma}
Algorithm \textnormal{Robust-$(P_2+P_4)$-Free-WED} is correct and runs in time $O(n m)$.
\label{lma:p2p4algo}
\end{lemma}
\begin{myproof}
\emph{Correctness:}
If $N_2 = \emptyset$, then $v$ is universal in $G$ and hence $D:=\left\{ v \right\}$.
If $G[R]$ contains an induced $P_4$, then together with $v$ and some vertex of $N_1$, $G$ contains an induced $P_2+P_4$.
Hence, Step (b) is correct.
Step (c) is correct by (\ref{eqn:independentU}).
The \texttt{Stop and Return} in Step (d) is correct, because in that case, the vertices of $N_2$ cannot be dominated, so there is no need to continue the search for a solution.
Step (f) is correct by (\ref{eqn:k1}).
Note that in Step (f) $U_1'$ can be empty and then $D$ is not an e.d.
But this is correct, because by (\ref{eqn:k1}) $G$ has no e.d.\ that contains $v$ in that case.
Step (g) is correct by (\ref{eqn:onemiss}).
The condition $|Z|=k$ in Step (i) implies by (\ref{eqn:onemiss}) that $|U_i \cap N_4| = 1$ for all $i \in \left\{ 1,\dots,j \right\}$, because otherwise the algorithm had stopped in Step (g).
Hence, Step (i) is correct by (\ref{eqn:alln4_1}) and (\ref{eqn:alln4_2}).
Analogously, the condition $|Z|=k-1$ in Step (j) implies that there is exactly one $i \in \left\{ 1,\dots,k \right\}$ such that $U_i \cap N_4 = \emptyset$.
The condition $m(y) = |N_2|$ clearly implies $N_2 \subseteq N(y)$.
Note that there may be no vertex $y \in U_i$ with $N_2 \subset N(y)$.
In that case, $D$ is not an e.d.\ of $G$.
Either way, Step (j) is correct by (\ref{eqn:almostalln4_1}) and (\ref{eqn:almostalln4_2}).
Step (k) is justified by (\ref{eqn:oneseesall}), because $n_i(x)=|U_i|$ implies $U_i \subseteq N(x)$.
Note again that not necessarily every non-neighbor $y_i$ exists and if one does not exist, $D$ is not an e.d.\ of $G$.
Finally, Step (l) is correct by (\ref{eqn:differentmisses}).
Again, it may happen that one of the common non-neighbors $y_i$ does not exist, but in this case $G$ has no e.d.\ containing $v$.

\emph{Time bound:}
Testing a graph for being cograph and therefore Step (b) can be done in time $O(n+m)$ by \cite{BreCorHabPau2008,CorPerSte1985}.
The components of a graph can also be found in time $O(n+m)$.
For every component, the universal vertices can simply be identified by counting the number of neighbors in that component.
We can assume that every vertex is already labeled with $N_1,N_2,\dots,N_5$ and, hence, deciding if a vertex $x$ is in $N_i$ for some $i$ can be done in constant time.
With this, Step (c) can be done in time $O(n+m)+O(n)$, because for checking if $U \cap N_3 = \emptyset$ or $|Q \cap N_4| > 1$ it suffices to touch every vertex of $R$ at most twice.
We can label the vertices in Step (c) with the component they belong to and give them additionally a label $U$ if they belong to $U$.
Hence, we can decide in constant time for a vertex $v$ if $v \in U_i$ for some $i \in \left\{ 1,\dots,k \right\}$.
Step (e) can be done by considering every $y_i \in U_i$ for all $i \in \left\{ 1,\dots,k \right\}$ and its neighbors in $N_2$.
For every neighbor $x \in N_2$ of $y_i$, increase $n_i(x)$ and $m(y_i)$ by one.
Hence, Step (e) takes at most time $O(m)$.
In Step (f), the set $U'_1$ can be computed by checking $m(y)$ for every $y \in U_1$ and memorize the vertex with $m(y)=|N_2|$ and minimum $\omega(y)$.
This takes at most $O(|U_1|) = O(n)$ time.
Step (g) can be done in $O(n+m)$ time, by considering every vertex $x\in N_2$ and all numbers $n_i(x)$
(the number of which is not bigger than $|N(x)|$).
Set $Z$ and its components of Step (h) can be computed when $m$ is calculated in Step (e).
In Step (i),
we can find the minimum of $\omega(y_i)-\omega(z_i)$ for every vertex $y_i \in U_i$ and every $i \in \left\{ 1,\dots,k \right\}$ in time $O(|U_1|+\dots+|U_k|)=O(n)$.
In Step (j), clearly we can find the index $i \in \left\{ 1,\dots,k \right\}$ such that $U_i \cap Z = \emptyset$ in time $O(n)$.
Then it takes at most $O(|U_i|)=O(n)$ time to find a vertex $y \in U_i$ with $m(y)=|N_2|$ of minimum weight.
In Step (k), a vertex $x \in N_2$ with $n_i(x)=|U_i|$ can be identified in linear time, it one exists.
Then, its non-neighbors can be collected by testing if $x$ is adjacent to $y_j$ for every $y_j \in U_j$ and every $j \in \left\{ 1,\dots,k \right\}$.
That takes at most $O(|U_1|+\dots+|U_k|)=O(n)$ time.
A vertex $y \in U_i$ with $m(y_i)=|N_2|$ of minimum weight can also be found in linear time.
When reaching Step (l), we have to find the vertices $x,x',a,b,c$ and $d$ as stated in the algorithm.
If they do not exist, we can simply return $\emptyset$, because then $G$ has no e.d.\ containing $v$.
To find the vertices we follow the proof of (\ref{eqn:differentmisses}).
Generally, if the vertices of a subset $M$ of $V$ are labeled with $M$, we can find for a given vertex $w$ a neighbor or a non-neighbor in $M$ in time $O(m)$ by considering every edge of $G$ and check if one of the endpoints is $w$ and the other endpoint is labeled with $M$.
Finding two vertices of $w,w' \in N_2$ with different non-neighbors $y_1,y_1' \in U_1$ can be done by choosing a vertex $w \in N_2$ arbitrarily, then finding a non-neighbor $y_1$ of $w$ in $U_1$, then finding a neighbor $w'$ of $y_1$ in $N_2$ and finally finding a non-neighbor $y_1'$ of $w'$ in $U_1$.
Since the vertices of $N_2$ and $U_1$ are labeled with $N_2$, respectively $U_1$, this can be done in time $O(3m)=O(m)$.
Finding non-neighbors $y_2$ and $y_2'$ of $w$ and $w'$ in $U_2$ can also be done in time $O(m)$, because the vertices of $U_2$ are labeled with $U_2$.
If $y_2\not=y_2'$, we are done.
Otherwise, find a neighbor $w'' \in N_2$ of $y_2$ in time $O(m)$ and its non-neighbor $y_1''$ in $U_1$ in time $O(m)$.
Then it can be checked in constant time which vertex to choose for $x,x',a,b,c$ and $d$ to fulfill the conditions.
The non-neighbors $y_3,\dots,y_k$ of $x$ in $U_3,\dots,U_k$ can be found by considering every neighbor of $x$ and mark it with $N(x)$ and then considering every vertex $y_i \in U_i$ and check if $y_i$ is labeled with $N(x)$ for all $i \in \left\{ 3,\dots,k \right\}$.
This takes at most $O(m+n)$ time.
Hence, Step (l) can be done in time $O(n+m)$ as well.

Thus, the algorithm \texttt{Robust-$(P_2+P_4)$-Free-WED} takes $O(n m)$ time.
\end{myproof}

\section{A Polynomial Time Algorithm for the $2$-Bounded WED Problem}\label{2BED}

For a non-negative integer $k$, an e.d.\ $D$ in a graph $G$ is said to be {\it $k$-bounded}
if every vertex in $D$ has degree at most $k$ in $G$. For short, a $k$-bounded e.d.\ will also be referred to as a {\it $k$-b.e.d.}.
The task of the $k$-Bounded Weighted Efficient Domination ($k$-BWED) problem is to determine whether a given vertex-weighted graph $G$ admits a $k$--b.e.d., and if so, to compute one of minimum weight.
Clearly, a graph $G$ admits a $0$-b.e.d.\ if and only if it is edgeless. It is also straightforward to see that
$G$ admits a $1$-b.e.d.\ if and only if each connected component of $G$ is either $K_1$, $K_2$, or the vertices of degree $1$ in it form an ED set. Therefore, the $k$-BWED problem is solvable in linear time for $k\in \{0,1\}$.
On the other hand, since the ED problem is \NP-complete for graphs of maximum degree $3$, 
the $k$-BWED problem is \NP-complete for every $k\ge 3$.
In the rest of the section, we prove that the $k$-BWED problem is also solvable in polynomial time for $k = 2$,
thus determining the computational complexity status of the $k$-BWED problem for every value of $k$.

For convenience, let us formally state the $2$-BWED problem again:

\medskip
\begin{center}
\fbox{\parbox{0.85\linewidth}{\noindent
{\sc $2$-Bounded Weighted Efficient Domination ($2$-BWED)}\\[.8ex]
\begin{tabular*}{.9\textwidth}{rl}
{\em Instance:} & A graph $G=(V,E)$, vertex weights $\omega:V\to \mathbb{N}$.\\
{\em Task:} & Find a $2$-b.e.d.\ of minimum total weight,\\
& or determine that $G$ contains no $2$-b.e.d.
\end{tabular*}
}}
\end{center}

\begin{theorem}
The $2$-BWED problem is solvable in polynomial time.
\end{theorem}

\begin{myproof}
In what follows, we describe an algorithm for the $2$-BWED problem.
First, the algorithm computes the connected components of $G$.
If $G$ is not connected, it solves the problem
recursively on connected components and combines the solutions in the obvious way.

If $G$ is a cycle, then one of the following two cases occurs:
\begin{itemize}
  \item Either $n\not\equiv 0 \pmod 3$, in which case $G$ has no e.d.\ and thus also no $2$-b.e.d. The algorithm returns {\sc no}.
  \item Or $n\equiv 0\pmod 3$, in which case
$G$ has exactly three $2$-b.e.d.s, namely
$D_0$, $D_1$, $D_2$, where
\hbox{$D_j = \{v_i\mid 1\le i\le n\,,~~ i\equiv j\pmod 3\}$} for $j\in \{0,1,2\}$,
where $(v_1,\ldots, v_{n})$ is a cyclic order of the vertices of $G$.
In this case, the algorithm returns the set of minimum weight among $D_0$, $D_1$, $D_2$.
\end{itemize}

From now on, we assume that $G$ is connected but not a cycle.
For this case, we will develop a polynomial time algorithm for
the following generalization of the $2$-BWED problem:

\medskip
\noindent
{\em Instance:} A graph $G=(V,E)$, vertex weights $\omega:V\to \mathbb{N}$, a subset $X\subseteq V$ such that for all $x\in X$, it holds $d(x)\le 2$.\\
{\em Task:} Find an e.d.\ $D\subseteq X$ of minimum total weight, or determine that $G$ contains no e.d.\ contained in $X$.

\medskip
For readability reasons, we describe the steps of the algorithm in {\it italic type}. After each step where the algorithm returns something, we justify the correctness of the step (assuming inductively that the algorithm works correctly on smaller instances).
If $G$ contains an e.d.\ $D$ with $D\subseteq X$, then the output of the algorithm will be a minimum weight e.d.\ with $D\subseteq X$. Otherwise, the output will be {\sc no}.

\smallskip
{\it {\bf Step $1$.}
Let $Y = V\setminus X$. Delete from $G$ all edges in $Y$.
If $Y$ has an isolated vertex, return {\sc no}.}
\smallskip

The correctness of Step $1$ follows from two facts:
(1) we may assume that $Y$ is independent since
the edges completely within $Y$ cannot be used for dominating any vertices;
(2) if after removing from $G$ all the edges of $Y$, there exists an isolated vertex in $Y$, then such a vertex cannot be dominated by any vertex in $X$, hence $G$ is a no instance.

From now on, we assume that $Y$ is independent and every vertex in $Y$ has a neighbor in $X$.

\smallskip
\begin{sloppypar}
{\it {\bf Step $2$.}
Compute the connected components $Q_1,\ldots, Q_k$ of $G$.
If $k>1$, solve the problem recursively on connected components,
with inputs \hbox{$(Q_i, \omega_{|V(Q_i)}, X\cap Q_i)$}.
If every connected component of $G$ is a yes instance, then return the union of recursively computed sets, one for each connected component of $G$. Else, return {\sc no}.}
\end{sloppypar}
\smallskip

The correctness of the above step is obvious.

%
%

From now on, we assume that $G$ is connected but not a cycle. Since $G$ is not a cycle, every connected component of $G[X]$ is a path. Moreover, the internal vertices of these paths have no neighbors outside $X$ since they are of degree $2$ in $G$.

\smallskip
{\it {\bf Step $3$.} If $X = \emptyset$, then return {\sc $\emptyset$}.}
\smallskip

The correctness of Step $3$ follows from the fact that if $X = \emptyset$, then also $Y = \emptyset$, since every vertex in $Y$ has a neighbor in $X$.

From now on, we assume that $X$ is nonempty.

\smallskip
\begin{sloppypar}
{\it {\bf Step $4$.}
If $G[X]$ contains a path $P=(v_1,\dots,v_k)$ with $k \equiv 0\pmod 3$, then let
$G'=G-V(P)$. Run the algorithm recursively on $I = (G', \omega_{|V(G')}, X\cap V(G'))$. If $I$ is a no instance,
return {\sc no}.
Else, return the union of the recursively computed set and the set \hbox{$\{v_i\mid 2\le i\le k-1\,,~~i\equiv 2\pmod 3\}$}.}
\end{sloppypar}
\smallskip

To justify the correctness of Step $4$, suppose first that $(G,\omega,X)$ is a yes instance to the problem, and let $D$ be an optimal solution. Then, $D\subseteq X$; moreover, either $v_1\in D$ or $v_2\in D$ (since otherwise $v_1$ would not be dominated) but not both (since otherwise $D$ would not be independent).
If $v_1\in D$, then $D\cap V(P) = \{v_i\mid 1\le i\le k-2\,,~~i\equiv 1\pmod 3\}$, which implies that $v_k$ is not dominated, a contradiction. Hence, $v_2\in D$ and consequently $D\cap V(P) = \{v_i\mid 2\le i\le k-1\,,~~i\equiv 2\pmod 3\}$.
Hence, every vertex in $Y$ is dominated by a (unique) vertex of $D\setminus V(P)$, and consequently
the set $D\setminus V(P)$ is a feasible solution for $I$.
Conversely, if $D'$ is an optimal solution for $I$, then
it is straightforward to verify that the set
\hbox{$D'\cup\{v_i\mid 2\le i\le k-1\,,~~i\equiv 2\pmod 3\}$} is a feasible solution for $(G,\omega,X)$.

From now on, we assume that every path in $G[X]$ is of order $1$ or $2 \pmod 3$.

\smallskip
{\it {\bf Step $5$.}
Compute the set ${\cal P}_2$ of all paths in $G[X]$ of order $k \equiv 2\pmod 3$.
It there exists a path $P\in {\cal P}_2$ with at least $5$ vertices,
then let $G' = (G-\{v_2,v_3,\ldots, v_{k-1}\}) + v_1v_k$,
where $P =(v_1,\ldots, v_{k})$.
Let $\omega':V(G')\to\mathbb{N}$ be defined as
$$\omega'(v)= \left\{
               \begin{array}{ll}
                 \sum_{i = 0}^{(k-2)/3}\omega(v_{3i+1}), & \hbox{if $v = v_1$;} \\
                 \sum_{i = 0}^{(k-2)/3}\omega(v_{3i+2}), & \hbox{if $v = v_k$;}\\
                 \omega(v), & \hbox{if $v\not\in \{v_1,v_k\}$.}
                \end{array}
             \right.$$
Run the algorithm recursively on $I = (G',\omega',X\cap V(G'))$.
If $I$ is a no instance to the problem, return {\sc no}.
Else, return the set $D'\cup D_0$ where $D'$ is the recursively computed set,
and
        $$D_0 = \left\{
        \begin{array}{ll}
        \{v_i\mid 4\le i\le k-1\,,~~i\equiv 1\pmod 3\}, & \hbox{if $v_1\in D'$;} \\
        \{v_i\mid 2\le i\le k-3\,,~~i\equiv 1\pmod 3\}, & \hbox{if $v_k\in D'$.}
        \end{array}
        \right.$$}
\smallskip

To justify the correctness of Step $5$, let $P\in {\cal P}_2$ be a path as above.
Suppose first that $(G,\omega,X)$ is a yes instance to the problem, and let $D$ be an optimal solution.
Then, $D\subseteq X$; moreover, either $v_1\in D$ or $v_2\in D$, but not both.
If $v_1\in D$, then $D\cap V(P) = \{v_i\mid 1\le i\le k-1\,,~~i\equiv 1\pmod 3\}$.
Since the internal vertices of $P$ do not dominate vertices outside $P$,
the set $D' = D\setminus \{v_2,v_3,\ldots, v_{k-1}\}$ is a feasible solution of $I$, with
$\omega'(D') = \omega(D)$. (The edge $v_1v_k$ is necessary so that $v_k$ is dominated.)
A similar reasoning can be used in the case when $v_2\in D$.
Conversely, if $D'$ is an optimal solution for $I$, then the set $D = D'\cup D_0$
where $D_0$ is as specified in Step $5$, is a feasible solution for $(G,\omega,X)$
with $\omega(D) = \omega'(D')$.

From now on, we assume that every path in ${\cal P}_2$ contains exactly two vertices.

\smallskip
{\it {\bf Step $6$.} For every path $P\in {\cal P}_2$ such that
its endpoints have a common neighbor in $Y$,
delete an endpoint of $P$ with maximum weight from $G$,
and remove $P$ from ${\cal P}_2$.}
\smallskip

To justify the correctness of Step $6$, let $P\in {\cal P}_2$ be a path such that
its endpoints $x$ and $x'$ have a common neighbor in $Y$. Suppose that $x'$ was the deleted vertex.
If $D\subseteq X$ is a minimum weight e.d.\ in $G$ (among all e.d.'s contained in $X$)
such that $x'\in D$ then $(D\setminus \{x'\})\cup\{x\}$ is also an optimal solution.
On the other hand, every optimal solution for the reduced instance
is a feasible (and, by the choice of $x'$ also optimal) solution for the original instance.

From now on, we assume that for every path in ${\cal P}_2$, its endpoints
have no common neighbor in $Y$.

\smallskip
\begin{sloppypar}
{\it {\bf Step $7$.}
Compute the set ${\cal P}_1$ of all paths in $G[X]$ of order $k \equiv 1\pmod 3$.
If there exists a path $P\in {\cal P}_1$ with at least $4$ vertices, then let
$G' = G-\{v_2,v_3,\ldots, v_{k-1}\}$, where $P =(v_1,\ldots, v_{k})$. Run the algorithm recursively on
$I = (G', \omega_{|V(G')}, X\cap V(G'))$. If $I$ is a no instance, return {\sc no}.
Else, return the union of the recursively computed set and the set \hbox{$\{v_i\mid 4\le i\le k-3\,,~~i\equiv 1\pmod 3\}$}.}
\end{sloppypar}
\smallskip

To justify the correctness of Step $7$, let $P\in {\cal P}_1$ be a path as above.
Suppose first that $(G,\omega,X)$ is a yes instance to the problem, and let $D$ be an optimal solution.
Then, $D\subseteq X$; moreover, either $v_1\in D$ or $v_2\in D$, but not both.
If $v_2\in D$, then $D\cap V(P) = \{v_i\mid 2\le i\le k-3\,,~~i\equiv 2\pmod 3\}$, which implies that $v_k$ is not dominated, a contradiction.
Hence, $v_1\in D$ and consequently $D\cap V(P) = \{v_i\mid 1\le i\le k\,,~~i\equiv 1\pmod 3\}$.
Since the internal vertices of $P$ do not dominate vertices outside $P$,
the set $D\setminus \{v_2,v_3,\ldots, v_{k-1}\}$ is a feasible solution in $G'$.
Conversely, if $D'$ is an optimal solution for the reduced instance,
then it is straightforward to verify that the set \hbox{$D'\cup\{v_i\mid 4\le i\le k-3\,,~~i\equiv 1\pmod 3\}$} is a feasible solution for the original instance $(G,\omega,X)$.

From now on, we assume that every path in ${\cal P}_1$ contains a single vertex.

\smallskip
{\it {\bf Step $8$.} If $Y = \emptyset$ then return any set $\{x\}$ with $x\in X$ minimizing the value of $\omega(x)$.}
\smallskip

To justify the correctness of Step $8$, note that if $Y = \emptyset$ then the connectedness of $G$ and the assumptions made after Steps $5$ and $7$ imply that $G[X]$ consists of a single component with at most two vertices. The conclusion follows.

From now on, we assume that $Y\neq \emptyset$. Consequently, since $G$ is connected, every component of $G[X]$ has a neighbor in $Y$.

We say that a vertex of $G$ is {\it forced} if it is contained in every e.d.\ set $D$ with $D\subseteq X$.

\smallskip
{\it {\bf Step $9$.} Let $I= \cup_{P\in {\cal P}_1}V(P)$ denote the set of vertices contained in a path from ${\cal P}_1$.
If two vertices in $I$ have a common neighbor, return {\sc no}.}
\smallskip

The correctness of Step $9$ follows from the fact that every vertex in $I$ is forced.
Indeed, by the assumption after Step $7$, every path in ${\cal P}_1$ contains a single vertex. Hence, $I$ is an independent set and every vertex in $I$ is forced since it can only be dominated by itself.

From now on, we assume that no two vertices in $I$ have a common neighbor.

\smallskip
{\it {\bf Step $10$.} Compute the set $M= \cup_{P\in {\cal P}_2}V(P)$, the set of vertices contained in a path from ${\cal P}_2$.}
\smallskip

By the assumption after Step $5$, every path in ${\cal P}_2$ has exactly two vertices. Hence $M$ induces a matching in $G$.
For simplicity, we will refer to edges of paths in  ${\cal P}_2$ as {\it edges of $M$}.
By the assumption after Step $6$, no edge of $M$ is contained in a triangle in $G$.
Note that $X$ is the disjoint union $X = I\cup M$, and for every edge $e$ of $M$, every e.d.\ $D$ of $G$
with $D\subseteq X$
(if there is one) contains exactly one endpoint of $e$ (since otherwise neither of the two endpoints of $e$ would be dominated exactly once).

\smallskip
{\it {\bf Step $11$.} If there exists a vertex $v\in I$, then let $N_1$ and $N_2$ be the sets of vertices at distance $1$ and $2$ from $v$ in $G$, respectively.
If $N_2$ contains an edge, then return {\sc no}. Else, let $G' = G-(\{v\}\cup N_1\cup N_2)$ and run the algorithm recursively on the instance $I = (G', \omega_{|V(G')},X\cap V(G'))$. If $I$ is a no instance, return {\sc no}.
Else, return the set $D'\cup \{v\}$ where $D'$ is the recursively computed set.}
\smallskip

To justify the correctness of Step $11$, let $v$, $N_1$ and $N_2$ be as in Step $11$.
Suppose first that $(G,\omega,X)$ is a yes instance, and let $D$ be an optimal solution.
Then, since $v$ is forced, $v\in D$. Clearly, we have $D\cap (N_1\cup N_2) = \emptyset$.
Moreover, since $Y$ is independent and no two vertices in $I$ have a common neighbor, $N_2$ is a subset of $M$.
In particular, every vertex $b\in N_2$ is an endpoint of an edge $e$ in $M$, and hence can only be dominated in $D$
by the other endpoint, say $b'$, of $e$. This implies that $N_2$ is an independent set (since otherwise an edge of $M$ would be contained in $N_2$).
Hence, if the algorithm returns {\sc no} in Step $11$, then  $(G,\omega,X)$ is indeed a no instance.
Moreover, if $D\subseteq X$ is an e.d.\ set of $G$, then
the set $D\setminus \{v\}$ is clearly a feasible solution for the reduced instance.
Conversely, suppose that $N_2$ is an independent set and let $D'$ be an optimal solution for the
reduced instance $I$. Every vertex $b'\in M$ the unique neighbor of which in $M$ belongs to $N_2$ can only be dominated in $D'$ by itself, hence it is forced in $D'$. Consequently, the set $D'\cup \{v\}$ is a feasible solution for the original instance.

From now on, we assume that $I = \emptyset$.

\smallskip
{\it {\bf Step $12$.} Compute the multigraph $H$ with $V(H) = Y$ in which two distinct vertices $y$, $y'$ are connected by exactly $k$ edges, where $k$ is the number of edges $xy$ in $M$ with $N_G(x) \cup N_G(y) = \{y,y'\}$. }
\smallskip

The multigraph $H$, which may have multiple edges, will help us determining whether $(G,\omega,X)$ is a yes instance.
By the connectedness of $G$ and the fact that every vertex in $Y$ has a neighbor in $X$, multigraph $H$ is connected and
contains at least one edge. Moreover, there is a bijective correspondence between
the edges of $H$ and edges of $M$, in the sense that
every edge $e$ in $H$ is generated by a unique edge $e_M$ of $M$,
and conversely, since every vertex in $M$ has at most one $G$-neighbor in $Y$, every edge $e$ of $M$ generates exactly one edge $e_H$ of $H$.
Recall that an {\it orientation} of a multigraph $H'$ is a directed multigraph obtained from $H'$ by assigning to each of its edges one of the two possible orientations.
A {\it $1$-orientation} of $H$ is an orientation of $H'$ in which every vertex has out-degree
exactly $1$.

\begin{sloppypar}
\begin{clai}\label{orientations-2-BED}
Graph $G$ has an e.d.\ set $D$ with $D\subseteq X$
if and only if $H$ admits a $1$-orientation.
\end{clai}
\end{sloppypar}

\begin{subproof}
Suppose first that $G$ has an e.d.\ $D\subseteq X$.
Then, $|D\cap e| = 1$ for every edge of $M$.
We will now describe how to obtain a $1$-orientation of $H$.
For every edge $e = \{y,y'\}\in E(H)$,
let $x$ be the vertex of $e_M$ contained in $D$.
Orient $e$ from $y$ to $y'$ if $y$ is the neighbor of $x$ in $e$ (in the graph $G$),
and from $y'$ to $y$, otherwise.
Since $D$ is an e.d.\ in $G$,
each vertex $y\in Y = V(H)$ has exactly one neighbor in $D$.
Therefore, for each vertex $y$, exactly one of the edges of $H$ incident with $y$ will be oriented
away from $y$ in the above orientation, and this is indeed a $1$-orientation of $H$.

Conversely, suppose that $H$ admits a $1$-orientation $\tilde H$.
For every vertex $y\in V(H) = Y$, let $x(y)$ be the unique element of $e_M\cap N_G(y)$
where $e= \{y,y'\}$ is the unique edge of $H$ oriented away from $y$ in $\tilde H$.
Let $D = \{x(y)\,:\,y\in Y\}\cup\{x(e)\,:\,e\in L\}$.
By construction, $D\subseteq X$,
$D$  contains exactly one endpoint of each edge of $M$, and every vertex in $Y$ has exactly one neighbor in $D$.
Hence, $D$ is an e.d.\ of $G$ with $D\subseteq X$.
\end{subproof}

Claim~\ref{orientations-2-BED} and its algorithmic proof shows that every feasible solution $D\subseteq X$
corresponds to a $1$-orientation of $H$, and vice versa.
The next claim characterizes the cases when $H$ admits a $1$-orientation.

\begin{sloppypar}
\begin{clai}\label{orientations-unicyclic}
Multigraph $H$ admits a $1$-orientation if and only if
it has exactly one cycle.
If this is the case, then $H$ admits exactly two $1$-orientations.
\end{clai}
\end{sloppypar}

\begin{subproof}
First, suppose that $H$ admits a $1$-orientation $\tilde H$.
The digraph $\tilde H$ cannot be acyclic since otherwise it would contain a sink
(a vertex with out-degree $0$). Let $C$ be a directed cycle in $\tilde H$.
Then every vertex in $C$ has exactly one out-neighbor in $C$, and hence it has no out-neighbors
outside $C$. Moreover, if two vertices of $C$ are connected by an edge $e$ in $H$, then
an orientation of $e$ is in $C$, since otherwise a vertex of $C$ would have out-degree at least $2$ in $\tilde H$.
Since $H$ is connected, it can be proved by induction on $k\ge 1$ that
for every vertex $v\in S_k$, where $S_k$ denotes the set of vertices in $V(H)\setminus C$ at distance $k$ from $C$,
there exists a directed path from $v$ to $C$ in $\tilde H$, and the set $S_k$ is independent in $H$.
Therefore, $H$ has a unique cycle.

Conversely, suppose that $H$ has a unique cycle $C$. A $1$-orientation of $H$ can be obtained by
orienting the edges in $C$ in one of the two directions following the cycle, and orienting all the other edges of $H$
toward $C$. Moreover, these are clearly the only possible $1$-orientations of $H$.
\end{subproof}

Claim~\ref{orientations-unicyclic} and its constructive proof justify the following final step of the algorithm.

\smallskip
{\it {\bf Step $13$.}
If $H$ is a tree or $H$ has at least two cycles, then return {\sc no}.
Otherwise,
compute the two $1$-orientations
$\tilde H_1$ and $\tilde H_2$ of $H$
(as in the second part of the proof of Claim~\ref{orientations-unicyclic}).
For each $i\in\{1,2\}$, compute
an e.d.\ set $D_i\subseteq X$ (as in the second part of the proof of Claim~\ref{orientations-2-BED}).
If $\omega(D_1)\le \omega(D_2)$, then return $D_1$. Otherwise,
return $D_2$.}
\smallskip

The correctness of the algorithm outlined in Steps $1$--$13$ follows from the above discussion.
Clearly, the algorithm can be implemented so that it runs in polynomial time.
\end{myproof}

\section{\NP-Completeness of the ED Problem}\label{EDNPc}

Recall that the ED problem is known to be \NP-complete on planar bipartite graphs~\cite{LuTan2002} and planar graphs with maximum degree $3$ \cite{Kratochvil91,FelHoo2000}.

\begin{theorem}\label{EDplanarbipdeg3NPc}
For every $g\ge 3$, the ED problem is \NP-complete on planar bipartite graphs of maximum degree $3$ with girth at least $g$.
\end{theorem}

We reduce from the \textsc{One-In-Three 3SAT} problem: Given a Boolean formula $F$ as 3-CNF, decide if there is a satisfying truth assignment such that every clause of $F$ contains exactly one true literal.
This problem remains \NP-complete even for monotone formulas whose incidence graph $I(F)$ is planar (see \cite{Laroche1992,MooreRobson2001,MulzerRote2008}).

Let $F$ be a monotone, planar 3-CNF with variables $v_1,\dots,v_n$ and clauses $C_1,\dots,C_m$ and let $I(F)=(\cV\cup\cC, \cE)$ be its incidence graph.
Fix a planar embedding of $I(F)$. For all $i \in \left\{ 1,\dots,n \right\}$, $j \in \left\{ 1,\dots,m \right\}$, let $a(i,j)$ be the position of the clause $C_j$ in a clockwise ordering of all neighbors of $v_i$ beginning with the clause of smallest index, or undefined, if $v_i \not \in C_j$.

\newcommand{\ow}{\overline{w}}
\newcommand{\ov}{\overline{v}}
\newcommand{\ox}{\overline{x}}

We construct the reduction graph $G(F)$ by modifying $I(F)$.
Replace every variable vertex $v_i \in \cV$ by the path $P_i$ of $6m$ vertices defined as:
\[ P_i = (\ow_{i,1},v_{i,1},\ox_{i,1},w_{i,1},\ov_{i,1},x_{i,1},\dots,\ow_{i,m},v_{i,m},\ox_{i,m},w_{i,m},\ov_{i,m},x_{i,m})\,,\]
and replace every edge $v_iC_j \in \cE$ by the path $E_{i,j}$ of $6g+2$ vertices defined as:
\[ E_{i,j} =(\ov_{i,a(i,j)},x_{i,j}^1,\ow_{i,j}^1,v_{i,j}^1,\ox_{i,j}^1,w_{i,j}^1,\ov_{i,j}^1,\dots,x_{i,j}^g,\ow_{i,j}^g,v_{i,j}^g,\ox_{i,j}^g,w_{i,j}^g,\ov_{i,j}^g,C_j)\,.\]
By the use of $a(i,j)$, $G(F)$ remains planar and by construction, every vertex has at most $3$ neighbors.
Furthermore, $G(F)$ has girth at least $g$, because the inserted paths that substitute variables are clearly acyclic, and every edge of $I(F)$ is replaced by an induced path of length $6g+1$.
Finally, $G(F)$ admits a bipartition by taking the overlined vertices in one independent set and the other vertices in the other independent set.

We define
\[ V_i = \left\{ v_{i,j}, \ov_{i,j}, v_{i,j}^k, \ov_{i,j}^k : 1 \le j \le m, 1 \le k \le g \right\} \]
and analogously $W_i$ and $X_i$ for all $i \in \left\{ 1,\dots,n \right\}$.
Let $D$ be an e.d.\ of $G(F)$.
Then for all $i \in \left\{ 1,\dots,n \right\}$:
\begin{equation}
\text{either } V_i \subseteq D \text{ or } W_i \subseteq D,
\label{eqn:npcopy}
\end{equation}
\begin{myproof}
Let $i \in \left\{ 1,\dots,n \right\}$.
The vertex $\ow_{i,1}$ has only one neighbor, namely $v_{i,1}$.
Since $\ow_{i,1}$ must be dominated by $D$, either $\ow_{i,1} \in D$ or $v_{i,1} \in D$.

If $\ow_{i,1} \in D$, then $v_{i,1}, \ox_{i,1} \not \in D$, because $D$ is efficient.
Then the only way to dominate $\ox_{i,1}$ is $w_{i,1} \in D$, because it has no other neighbors.
Repeating this argumentation along the path $P_i$ yields $\ow_{i,2},w_{i,2},\dots,\ow_{i,m},w_{i,m} \in D$.
With this, we have $x_{i,j}^1 \not \in D$ for all $j \in \left\{ 1,\dots,m \right\}$.
Then the only way to dominate $x_{i,j}^1$ is $\ow_{i,j}^1 \in D$.
Then by similar argumentation we get
\[w_{i,j}^1,\ow_{i,j}^2,w_{i,j}^2,\dots,\ow_{i,j}^g,w_{i,j}^g \in D \text{ for all } j \in \left\{ 1,\dots,m \right\}.\]

If $v_{i,1} \in D$, then $\ox_{i,1}, w_{i,1} \not \in D$ and the only way to dominate $w_{i,1}$ is $\ov_{i,1} \in D$.
Analogous to the first case, this implies $v_{i,2},\ov_{i,2},\dots,v_{i,m},\ov_{i,m} \in D$ and
\[v_{i,j}^1,\ov_{i,j}^1,\dots,v_{i,j}^g,\ov_{i,j}^g \in D \text{ for all } j \in \left\{ 1,\dots,m \right\}.\]
\end{myproof}
Since $D$ is efficient, (\ref{eqn:npcopy}) implies
\begin{equation}
\left( X_1 \cup \dots \cup X_n \right) \cap D = \emptyset \text{ and } \left\{ C_1,\dots,C_m \right\} \cap D = \emptyset.
\label{eqn:npwelldefined}
\end{equation}
Hence, each clause vertex is dominated by exactly one of its neighbors.
This means that if $G(F)$ admits an e.d.\ $D$, then $F$ is satisfied by the truth assignment that sets a variable $v_i$ true if and only if $V_i \in D$.

For the other direction, if $F$ is satisfiable by a truth assignment, let $D$ be the set containing all vertices in $V_i$ for every true variable $v_i$ and all vertices in $W_i$ for every false variable $v_i$, and nothing else.
Clearly, $D$ fulfills (\ref{eqn:npcopy}) and (\ref{eqn:npwelldefined}) and since the truth assignment sets exactly one variable per clause to true, every clause vertex of $G(F)$ has exactly one neighbor in $D$.
Hence, $D$ is an e.d.\ of $G(F)$.

Since $G(F)$ can be constructed in polynomial time, this proves Theorem \ref{EDplanarbipdeg3NPc}.

\begin{footnotesize}

\end{footnotesize}


\begin{thebibliography}{99}

\bibitem{BanBarSla1988}
    D.W. Bange, A.E. Barkauskas, P.J. Slater,
    Efficient dominating sets in graphs,
    in: R.D. Ringeisen and F.S. Roberts, eds., Applications of Discrete Math. (SIAM, Philadelphia, 1988) 189--199.

\bibitem{BanBarHosSla1996}
    D.W. Bange, A.E. Barkauskas, L.H. Host, P.J. Slater,
    Generalized domination and efficient domination in graphs,
    {\sl Discrete Math.} 159 (1996) 1-11.

\bibitem{Biggs1973}
   N.~Biggs,
   Perfect codes in graphs,
   {\sl J. of Combinatorial Theory (B)}, 15 (1973) 289-296.

\bibitem{BraHunNev2010}
    A. Brandst\"adt, C. Hundt, R. Nevries,
    Efficient Edge Domination on Hole-Free graphs in Polynomial Time,
    Conference Proceedings LATIN 2010, LNCS 6034, (2010) 650-661.

\bibitem{BraLeiRau2012}
    A. Brandst\"adt, A. Leitert, D. Rautenbach,
    Efficient Dominating and Edge Dominating Sets for Graphs and Hypergraphs,
    extended abstract in: Conference Proceedings of ISAAC 2012, LNCS 7676, 2012, 267-277.

\bibitem{BraMos2011}
    A. Brandst\"adt, R. Mosca,
    Dominating induced matchings for $P_7$-free graphs in linear time,
    extended abstract in: T. Asano et al. (Eds.): Proceedings of ISAAC 2011, LNCS 7074, 100-109, 2011. Available online in {\sl Algorithmica} 2012.

\bibitem{BreCorHabPau2008}
    A. Bretscher, D.G. Corneil, M. Habib, Ch. Paul,
    A Simple Linear Time LexBFS Cograph Recognition Algorithm,
    {\sl SIAM J. Discrete Math.} 22(4) (2008) 1277-1296.

\bibitem{CarKorLoz2011}
     D.M.~Cardozo, N. Korpelainen, V.V.~Lozin,
     On the complexity of the dominating induced matching problem in hereditary classes of graphs,
     {\sl Discrete Applied Math.} 159 (2011) 521-531.

\bibitem{C97}
M.-S. Chang, Weighted domination of cocomparability graphs, {\sl Discrete Appl. Math.} 80 (1997)
135-148.

\bibitem{ChaLiu1993}
    M.-S. Chang, Y.-C. Liu,
    Polynomial algorithms for the weighted perfect domination problems on chordal graphs and split graphs,
    {\sl Information Processing Letters} 48 (1993) 205-210.

\bibitem{CL94}
M-S. Chang and Y.-C. Liu,
Polynomial algorithms for weighted perfect domination problems on interval and circular-arc graphs,
{\sl J. Inf. Sci. Eng. 11} (1994) 549-568.

\bibitem{ChaPanCoo1995}
    G.J. Chang, C. Pandu Rangan, S.R. Coorg,
    Weighted independent perfect domination on co-comparability graphs,
    {\sl Discrete Applied Math.} 63 (1995) 215-222.

\bibitem{CorPerSte1985}
    D.G. Corneil, Y. Perl, L.K. Stewart,
    A linear recognition algorithm for cographs,
    {\sl SIAM J. Computing} 14 (1985) 926-934.

\bibitem{CouMakRot2000}
   B. Courcelle, J.A. Makowsky and U. Rotics,
   Linear time solvable optimization problems on graphs of bounded clique width,
   {\sl Theory of Computing Systems} {\bf 33} (2000) 125-150.

\bibitem{FelHoo2000}
   M.R. Fellows, M.N. Hoover,
   Perfect Domination,
   {\sl Australasian J. of Combinatorics} 3 (1991) 141-150.

\bibitem{FoeHam1977}
    S. F\"oldes, P.L. Hammer,
    Split graphs,
    {\sl Congressus Numerantium} 19 (1977) 311-315.

\bibitem{GarJoh1979}
    M.R. Garey, D.S. Johnson,
    Computers and Intractability -- A Guide to the Theory of NP-completeness,
    {\sl Freeman}, San Francisco, 1979.

\bibitem{GriSlaSheHol1993}
    D.L. Grinstead, P.L. Slater, N.A. Sherwani, N.D. Holmes,
    Efficient edge domination problems in graphs,
    {\sl Information Processing Letters} 48 (1993) 221-228.

\bibitem{Karp1972}
    R.M. Karp,
    Reducibility among combinatorial problems,
    In: {\em Complexity of Computer Computations}, Plenum Press, New York (1972) 85-103.

\bibitem{Kratochvil91}
J. Kratochv\'il, Perfect codes in general graphs, Rozpravy \v Ceskoslovensk\'e Akad. V\v ed \v Rada
Mat. P\v r\'irod V\v d 7 (Akademia, Praha, 1991).

\bibitem{Laroche1992}
 P. Laroche,
 Planar 1-in-3 satisfiability is NP-complete,
 ASMICS Workshop on Tilings, Deuxi\`eme Journ\'ees Polyominos et pavages,
 Ecole Normale Sup\'erieure de Lyon, 1992.

\bibitem{Leite2012}
    A. Leitert,
    Das Dominating Induced Matching Problem f\"ur azyklische Hypergraphen,
    Diploma Thesis, University of Rostock, Germany, 2012.

\bibitem{LiaLuTan1997}
   Y.D. Liang, C.L. Lu, C.Y. Tang,
   Efficient domination on permutation graphs and trapezoid graphs,
   in: {\sl Proceedings COCOON'97}, T. Jiang and D.T. Lee, eds., {\sl Lecture Notes in Computer Science} Vol. 1276
   (1997) 232-241.

\bibitem{Lin1998}
   Y.-L. Lin,
   Fast algorithms for independent domination and efficient domination in trapezoid graphs,
   in: {\sl Proceedings ISAAC'98}, {\sl Lecture Notes in Computer Science} Vol. 1533
   (1998) 267-275.

\bibitem{LivSto1988}
   M. Livingston, Q. Stout,
   Distributing resources in hypercube computers,
   in: {\sl Proceedings 3rd Conf. on Hypercube Concurrent Computers and Applications} (1988) 222-231.

\bibitem{LozMos2009}
    V.V. Lozin, R. Mosca,
    Maximum independent sets in subclasses of $P_5$-free graphs,
    {\sl Information Processing Letters} 109 (2009) 319-324.


\bibitem{LuTan1998}
    C.L. Lu, C.Y. Tang, Solving the weighted efficient edge domination problem on bipartite permutation graphs,
   {\sl Discrete Applied Math.} 87 (1998) 203-211.

\bibitem{LuTan2002}
    C.L. Lu, C.Y. Tang, Weighted efficient domination problem on some perfect graphs,
   {\sl Discrete Applied Math.} 117 (2002) 163--182.

\bibitem{LuKoTan2002}
    C.L. Lu, M.-T. Ko, C.Y. Tang,
    Perfect edge domination and efficient edge domination in graphs,
{\sl Discrete Applied Math.} 119 (2002) 227-250.

\bibitem{McCSpi1999}
    R.M. McConnell, J.P. Spinrad,
    Modular decomposition and transitive orientation,
    {\sl Discrete Math.} 201 (1999) 189-241.

\bibitem{Milan2012}
    M. Milani\v c,
    Hereditary Efficiently Dominatable Graphs,
    available online in: {\sl Journal of Graph Theory} 2012.

\bibitem{MooreRobson2001}
 C. Moore, J.M. Robson,
 Hard Tiling Problems with Simple Tiles,
 \emph{Discrete \& Computational Geometry} 26 (2001) 573--590.

\bibitem{MulzerRote2008}
 W. Mulzer, G. Rote,
 Minimun-Weight Triangulation is NP-hard,
 \emph{J. ACM} 55 (2008), Article No. 11.

\bibitem{RanSch2010}
  B. Randerath, I. Schiermeyer,
  On maximum independent sets in $P_5$-free graphs,
  \textsl{Discrete Applied Mathematics} 158 (2010), 1041-1044.

\bibitem{Spinr2003}
    J.P. Spinrad,
    Efficient Graph Representations,
    Fields Institute Monographs, American Math. Society, 2003.

\bibitem{Yen1992}
    C.-C. Yen,
    Algorithmic aspects of perfect domination,
    Ph.D. Thesis, Institute of Information Science, National Tsing Hua University, Taiwan 1992.

\bibitem{YenLee1996}
    C.-C. Yen, R.C.T. Lee,
    The weighted perfect domination problem and its variants,
{\sl Discrete Applied Math.} 66 (1996) 147-160.

\end{thebibliography}
\end{document}